\documentclass[pdflatex,sn-nature,iicol]{sn-jnl}
\usepackage{fix-cm}
\usepackage{graphicx}%
\usepackage{multirow}%
\usepackage[utf8]{inputenc}
\usepackage[T1]{fontenc}
\usepackage[english]{babel}
\usepackage{hyphenat}
\usepackage{amssymb,amsfonts}%
\usepackage[cal=cm]{mathalpha}
\usepackage{times}
\usepackage{mathptmx}
\usepackage{cuted}
\usepackage{mathtools}
\usepackage[table]{xcolor}%
\usepackage{textcomp}%
\usepackage{manyfoot}%
\usepackage{booktabs}%
\usepackage{listings}%
\usepackage[document, bottom]{ragged2e}
\usepackage{slashed}
\usepackage{setspace}
\usepackage{enumerate,enumitem}
\usepackage{wrapfig}
\usepackage{caption}
\usepackage{makecell}
\usepackage{centeredline}
\usepackage[title]{appendix}%
\usepackage{afterpage}
\usepackage[bottom]{footmisc}
\newcommand\scalemath[2]{\scalebox{#1}{\mbox{\ensuremath{\displaystyle #2}}}} % for equation scalling
\newcommand{\abs}[1]{\left| #1 \right|} % for absolute value
\newcommand{\avg}[1]{\left< #1 \right>} % for average
\newcommand{\ket}[1]{\left| #1 \right>} % for Dirac bras
\newcommand{\bra}[1]{\left< #1 \right|} % for Dirac kets
\newcommand{\braket}[2]{\left< #1 \vphantom{#2} \right|
	\left. #2 \vphantom{#1} \right>} % for Dirac brackets
\let\baraccent=\= % rename builtin command \= to \baraccent
\renewcommand{\=}[1]{\stackrel{#1}{=}} % for putting numbers above = 
\DeclareSymbolFont{eulerup}{U}{zeur}{m}{n}
\DeclareMathSymbol{\upgamma}{\mathord}{eulerup}{"0D} % to differentiate Euler-Mascheroni constant from gamma matrices
\newcommand{\olsi}[1]{\,\overline{\!{#1}}} % for better looking conjugated spinors \bar{\psi}

\begin{document}
	\captionsetup[figure]{font=small}
	\captionsetup[table]{font=small}
	%\onehalfspacing
	\justifying
	\raggedbottom
	%Title of paper Euler--Heisenberg
	
	\title[Euler--Heisenberg actions for gauge-axial background vectors]{Euler--Heisenberg actions for gauge-axial background vectors}
	\subtitle{Hessian diagonalization, sector classification, and applications\thanks{This work is based on the author's M.Sc. dissertation (Ref. \cite{mydissertation}), but further correcting and expanding upon it.}}
	%
	%\author{Lucas Pereira de Souza\inst{1}}
	%\thanks{This work is based on the author's M.Sc. dissertation (Ref. \cite{mydissertation}), but further correcting and expanding upon it.} %\emph{Present address:} Insert the address here if needed
	%\email[]{} 
	%for more information, reach out via e-mail: lucas.pereira@estudante.ufjf.br}
%\institute{Departamento de Física- ICE/UFJF, Via Local, 880. Rua José Lourenço Kelmer, São Pedro, Juiz de Fora-- MG, Brazil, CEP: 36036-900. \email{lucas.pereira@estudante.ufjf.br}}
%
\date{Received: \today / Revised version: \today}


\author*[1]{\fnm{Lucas} \sur{Pereira de Souza}}\email{lucas.pereira@estudante.ufjf.br}

\affil*[1]{\orgdiv{Departamento de Física - ICE}, \orgname{UFJF}, \orgaddress{\street{Via Local, 880. Rua José Lourenço Kelmer - São Pedro}, \city{Juiz de Fora}, \postcode{36036-900}, \state{Minas Gerais}, \country{Brazil}}}

\abstract{%
	%\vskip 6mm
	%
	%\centeredline{\textbf{\large Abstract}}
	%
	%\vskip 3mm
	%
	\justifying
	\noindent
	We derive the closed-form one-loop Euler--Heisenberg effective actions for Dirac fermions coupled simultaneously to classical electromagnetic vector and massive pseudo-vector backgrounds within a controlled quasi-static approximation. Through complete diagonalization of the functional Hessian, we systematically delineate the parameter space into distinct sectors characterized by stability properties and spectral structure. We identify subspaces that encompass and extend results from previous studies into a broader class, admitting propagating axial fields as physically viable regimes; strikingly, we note sectors presenting chirality-asymmetric instability. This addresses long-standing questions regarding the well-defined nature, diagonalizability, and stability of these models. From the effective actions, we derive novel nonperturbative pair-production rates for simultaneously propagating electromagnetic and axial vector backgrounds; remarkably, we find pronounced vacuum stabilization compared to previous results. Furthermore, we derive the non-perturbative contributions to the chiral current divergence, which enables dynamical chiral symmetry breaking and show that the electromagnetic coupling induces instanton-like configurations for the axial field, even when it is not a fundamental gauge field. As a proof-of-concept, we analyze a cosmological toy model of baryogenesis driven by an axial vector, providing numerical estimates that support the viability of this hypothesis. Additionally, we outline qualitative predictions for Weyl/Dirac semi-metals and briefly discuss potential applications in related phenomena, such as Quark-Gluon plasma, Electroweak scenarios and the Strong-CP problem.
	%
	%\PACS{
		%	{12.20.-m}{discribing text of that key}   \and
		%	{11.30.Rd}{discribing text of that key}   \and
		%	{11.10.Gh}{discribing text of that key}   \and
		%	{72.80.Vp}{discribing text of that key}   \and
		%} % end of PACS codes
}

\maketitle

%\tableofcontents

\section{Introduction\label{sec:intro}}

The nonperturbative response of the quantum vacuum to strong classical fields is a cornerstone of quantum electrodynamics (QED). Integrating out charged fermions yields the Euler--Heisenberg (EH) effective action \cite{EHoriginal}, which encodes low-energy ($\leq m c^2$) nonlinear photon interactions and underpins quantitative predictions for Schwinger pair production  \cite{schwinger1951}, vacuum birefringence, photon-photon scattering \cite{LASERIX,BIREF-HIBEF,CoRELS}, and related phenomena some of which are currently under experimental investigation. For homogeneous electromagnetic and/or scalar backgrounds, closed-form EH expressions are routinely derived via resummation \cite{HeatKernel1,HeatKernel2,barvinsky-vilkovisky,resummedeffectiveaction,sebastianresummedheatkerneleffective} or worldline \cite{numerical-wordline} methods, establishing the standard framework for field-induced particle production and nonlinear optics \cite{DunneEH,exactplanewave,exactselfdual1,exactselfdual2,exactsauter,exactplanewave2,nonlocalEffActionSebastian}. In recent decades, this formalism has been extended to include fermions coupled simultaneously to a vector potential $A_\mu$ and a parity-violating axial-vector field $S_\mu$---a configuration that arises in three distinct physical arenas: (i) gravitational theories with torsion, where the pseudo-trace of the contorsion tensor acts as $S_\mu$ and can seed baryogenesis \cite{HehlEMcoupling,stringtorsion,artigodoprof2002,ShapiroBelyaevTorsion,baryogenesismagnetic,KalbRamond,2Bornot2B}; (ii) high-energy extensions of the Standard Model, where axial fields may catalyze magnetic monopole pair production \cite{rajantiemonopole,monopole}; and (iii) Weyl / Dirac semimetals, where momentum-space node separations modulated by strain or magnetic textures generate emergent axial couplings \cite{semimetalsanomaly2,torsionweylmetal,Topologicalmaterials}. Despite their disparate origins, these contexts share a common field-theoretic core, motivating a unified nonperturbative treatment capable of bridging cosmological dynamics and condensed matter experiments.

Pioneering work by Maroto \cite{MarotoParticleProd,maroto99} first revealed that constant axial backgrounds in quasi-static configurations ($\partial_\mu S^\mu = 0$) can suppress Schwinger pair production---a stark departure from pure QED. Subsequent studies generalized the EH action to vector-axial systems \cite{exactaxialother,LorentzbreakingEHPetrov,bastianelliunifiedworldlineaxial,LorentzbreakingEHtensorPetrov}, but often yielded operator-valued expressions requiring further contractions, limiting practical utility. Recent advances identified special field configurations---e.g., constant, light-like, or parallel $S_\mu$---that admit exact diagonalization and exhibit either enhanced or suppressed pair creation \cite{maroto99,exactaxial,Copinger1,Sebastianheatkernelsresummationsspinor,CopingerSchwinger}. Concurrently, condensed matter theory has clarified how strain and Floquet driving emulate axial fields in topological semimetals, opening pathways to test Lorentz violation and nonlinear axial responses in lab \cite{EHinDiraccondensedmatter,Diractorsion2,Diractorsion,nonperturbLorentz,Lorentzviolationsemimetals2}. Yet a systematic classification of the full parameter space---delineating when the model is stable, Hermitian, and diagonalizable---has remained absent. The core obstacle lies in the functional Hessian: axial-vector interference can render it non-Hermitian or defective, signaling vacuum instability. Moreover, consistency conditions from torsion-motivated UV completions (e.g., Planck-scale mass, ghost-freedom \cite{torsionconsistency}) further constrain viable regimes, demanding a rigorous sector-by-sector analysis of physical admissibility.

In this work, we present a complete one-loop analysis of models containing fermions simultaneously coupled to a vector and a possibly massive pseudo-vector fields in configurations with constant field invariants (quasi-static/adiabatic settings), which we shall call electroaxial theories in what follows. Our work builds upon and extends the foundations laid before through the Following contributions: (i) The \emph{Complete Hessian diagonalization and systematic parameter-space classification} of well-defined and stable subspaces; (ii) The derivation of \emph{Exact closed-form effective actions} within the viable sectors even for propagating axial-vector field; (iii) The \emph{derivation of the chiral current divergence's structure} and demonstration of constraints that forces the pseudo-vector to \emph{inherit topological structures} from the gauge vector field; (iv) The extraction of \emph{novel Pair-production rates} for propagating axial-vector backgrounds and (v) The \emph{Phenomenological applications} with numerical estimates for relic field strength from a proof-of-concept toy-model for baryogenesis and qualitative assessments for condensed matter and other possible applications. Remarkably, our results indicate the possibility of chirality-asymmetric instability; a pronounced vacuum-stabilization when the pseudo-vector is present and estimate bounds for the viability of hypotheses in which baryogenesis is driven by an axial-vector.

The remainder of this paper is organized as follows. Section \ref{sec:electroaxial} reviews the theoretical framework: the electroaxial action and proper-time representation. Section \ref{sec:hessian} presents the complete Hessian diagonalization and parameter-space classification, establishing stability criteria for each sector. Section \ref{sec:exact} derives the exact one-loop effective action in closed form through zeta-function regularization, with detailed consistency checks against perturbative expansions and known limits. Section \ref{sec:Anom} analyzes the anomaly structure, the dynamical contributions to the divergence of the chiral current and topological constraints on axial vector configurations. Section \ref{sec:Apps} extracts vacuum-transition-probability measures and pair-production rates, including the new results for propagating axial fields and explores phenomenological implications: cosmological baryogenesis (Sec. \ref{subsec:cosmology}) and further applications (Sec. \ref{subsec:other}) with qualitative predictions for condensed matter and brief remarks on connections to the strong CP problem, dark energy, and current experimental searches for magnetic monopoles. We conclude in Section \ref{sec:conclusions}.

Throughout, we employ natural units ($\hbar = c = 1$), the Minkowski metric signature $(+,-,-,-)$, and adopt conventions consistent with Refs. \cite{shapiroprimer,schwartz,QFTQGdoprof} for spinor algebra. Numerical estimates use $M_{\mathrm{Planck}} = 1.22 \times 10^{19}\,\mathrm{GeV}$ and Planck 2018 cosmological parameters \cite{PlanckColab}.
%\vfill
%\pagebreak

\section{The Electroaxial Theory\label{sec:electroaxial}}

The strategy adopted throughout closely follows the approach established by Maroto \cite{maroto99}, which is itself an extension of the seminal tratment of pure QED inaugurated by Schwinger \cite{schwinger1951}---a procedure systematically reviewed in Refs. \cite{Dobadolivro,DunneEH}. We assume a familiarity with functional methods at the level of standard quantum field theory texts \cite{schwartz,QFTQGdoprof} while providing sufficient detail to make the calculations self-contained and establish notation conventions consistent with the broader literature on background-field effective actions.

We analyze a model-dependent theory of Dirac fermions $\psi$ coupled concurrently to an electromagnetic vector potential $A_\mu$ and a (possibly massive) pseudo-vector field $S_\mu$. The classical action in Minkowski spacetime is expressed as:
\begin{equation}
	\label{eq:Full_act}
	\mathrm{S} \left[A, S,\olsi{\psi},\psi\right] = \mathrm{S}_0 \left[A, S\right] + \mathrm{S}_{\text{Dirac}}.
\end{equation}
Here, $\mathrm{S}_0\left[A, S\right]$ pertains only to the vector fields, incorporating the standard Maxwell action:
\begin{equation}
	\label{eq:Maxwell_act}
	\mathrm{S}_{\text{Maxwell}} =-\frac{1}{4}\int d^4x \, F_{\mu\nu}F^{\mu\nu}
\end{equation}
along with the free axial-vector action \cite{artigodoprof2002,ShapiroBelyaevTorsion}:
\begin{equation}
	\label{eq:Proca_act}
	\mathrm{S}_{\text{Proca}} = \int d^4 x \left[-\frac{1}{4} S_{\mu\nu} S^{\mu\nu} + \frac{1}{2} M_S^2 S_\mu S^\mu\right].
\end{equation}
In these expressions, $F_{\mu\nu} = \partial_\mu A_\nu- \partial_\nu A_\mu$ and $S_{\mu\nu} = \partial_\mu S_\nu- \partial_\nu S_\mu$ are the field-strength tensors corresponding to the vector and pseudo-vector sectors, respectively, while $M_S$ denotes the mass of the pseudo-vector field. Based on the model of choice, classical interaction terms (e.g., $A_\mu S^\mu$) can also be included in the analysis. 

The Dirac action for the fermions is defined as: 
\begin{equation}
	\label{eq:Dirac_act}
	\mathrm{S}_{\text{Dirac}} = \int d^4x \, \olsi{\psi}(i \slashed{D}- m)\psi,
\end{equation}
where the covariant derivative is expressed as $\slashed{D} = \gamma^\mu (\partial_\mu- i e A_\mu- i \eta S_\mu \gamma^5)$. Here, $e$ and $\eta$ are the respective coupling constants to the vector and axial-vector backgrounds, while $\gamma^5 = i\gamma^0\gamma^1\gamma^2\gamma^3$ satisfies $\{\gamma^5, \gamma^\mu\} = 0$. The presence of $\gamma^5$ distinguishes this theory qualitatively from standard QED: left- and right-handed fermion components, obtainable through the application of the chiral projectors $\mathrm{P}_{L/R} = (1 \mp \gamma^5)/2$, couple to different ``effective vectors'':
\begin{equation}
	\label{eq:Vec_comb}
	V_\mu^{(R/L)} = e A_\mu \pm \eta S_\mu
\end{equation}
rather than to a common gauge field. This allows for the convenient notation $D_\mu = \partial_\mu-i V^{(R)}_\mu \mathrm{P}_{R}-i V^{(L)}_\mu \mathrm{P}_{L}$ for the covariant derivative.

The effective action can then be constructed by integrating out the fermions within the generating functional:
\begin{equation}
	\label{eq:gen_func}
	Z \left[A, S\right] = \frac{\int \mathcal{D} \olsi{\psi} \mathcal{D} \psi e^{i \mathrm{S} \left[A, S,\olsi{\psi},\psi\right]}}{\int \mathcal{D} \olsi{\psi} \mathcal{D} \psi e^{i \mathrm{S} \left[A = S = 0 \right]}} = e^{i W \left[A, S\right]},
\end{equation}
yielding the effective action:
\begin{equation}
	\label{eq:eff_action}
	W \left[A, S\right] = \mathrm{S}_0 \left[A, S\right]-i \mathrm{Tr} \ \mathrm{Ln} {\left(\frac{i\slashed{D}-m}{i\slashed{\partial}-m}\right)}.
\end{equation}
The one-loop quantum corrections are encapsulated in the last term, which forms the focal point of our analysis.

For simplicity, we assume that the background fields are smooth and confined to a compact spatial domain, with appropriate boundary conditions, within which the quasi-static approximation holds \cite{staticproca}. This assumption is standard for EH-type effective actions \cite{schwinger1951,DunneEH} as it enables the application of translation operator manipulations described in Sec. \ref{sec:exact} while effectively capturing the physics of slowly varying fields. Generalizations to time-dependent backgrounds extend beyond one-loop exactness and typically require resummation techniques \cite{HeatKernel1,HeatKernel2,barvinsky-vilkovisky,resummedeffectiveaction,sebastianresummedheatkerneleffective} or worldline methods \cite{numerical-wordline}. Under this quasi-static approximation, the extended Dirac operator exhibits elliptic properties and possesses a discrete spectrum, thereby ensuring it is trace-class \cite{HeatKernel1,HeatKernel2}. Consequently, the traces of $\slashed{D}$ and its transpose are equivalent, such that
\begin{equation}
	\label{eq:trlnequiv}
	\mathrm{Tr} \ \mathrm{Ln} \left(\frac{i\slashed{D}-m}{i\slashed{\partial}-m}\right) = \mathrm{Tr} \ \mathrm{Ln} \left(\frac{i\slashed{D}^\mathrm{T}-m}{i\slashed{\partial}^\mathrm{T}-m} \right).
\end{equation}
The transpose of the Dirac operator can be derived using the charge conjugation matrix defined as $C = i \gamma^2 \gamma^0$. This matrix exhibits the properties:
\begin{equation}
	\label{eq:C_prop}
	\begin{aligned}
		\scalemath{0.98}{C \gamma^\mu C^{-1} =-\left(\gamma^{\mu} \right)^\mathrm{T} \ \ \text{and} \ \ C \gamma^5 C^{-1} = \left(\gamma^5 \right)^\mathrm{T} = \gamma^5}.
	\end{aligned}
\end{equation}
These properties yield:
\begin{equation}
	\label{eq:transposeD}
	\hspace{-5pt}
	\scalemath{0.9}{\slashed{D}^\mathrm{T} =-C \left(\slashed{\partial}-i \slashed{V}^{(R)} \mathrm{P}_{L}-i \slashed{V}^{(L)} \mathrm{P}_{R}\right) C^{-1} =-C \slashed{D}^{\ast} C^{-1},}
\end{equation}
where $\slashed{D}^{\ast} = \slashed{\partial}-i e \slashed{A} + i \eta \slashed{S} \gamma^5$. Consequently, we can express the quantum correction as:
\vspace{-10pt}
\begin{equation}
	\label{eq:trlnequivr1}
	\hspace{-9pt}
	\begin{aligned}
		\scalemath{0.82}{\mathrm{Tr} \, \ \mathrm{Ln} \left(\frac{i\slashed{D}-m}{i\slashed{\partial}-m}\right)} & = \scalemath{0.82}{\mathrm{Tr} \ \mathrm{Ln} {\left[\frac{1}{\left(i\slashed{\partial}^\mathrm{T}-m\right)} C C^{-1} \left(i\slashed{D}^\mathrm{T}-m\right) \right]}}, \\
		& \scalemath{0.82}{= \mathrm{Tr} \ \mathrm{Ln} {\left[\frac{ \left(-i \slashed{D}^{\ast}-m \right)}{\left(-i \slashed{\partial}-m\right)} \right]}}.
	\end{aligned}
\end{equation}
%\vspace{-10pt}
Adding the term on the left-hand side, we find that this correction can be represented as:\\
\begin{equation}
	\label{eq:trlnequiv2}
	\scalemath{0.97}{i \overline{\Gamma} \left[A, S\right] = \frac{1}{2} \mathrm{Tr} \ \mathrm{Ln} {\left[\frac{{\slashed{D}^{\ast}} \slashed{D} - i m \left(\slashed{D}-\slashed{D}^{\ast} \right) + m^2}{\slashed{\partial}^2 + m^2}\right]}}.
	%\vspace{-10pt}
\end{equation}
\\In this equation, we denote the quantum corrections as $\overline{\Gamma} [A, S]$. Direct substitution gives:\\
%\vspace{0pt}
\begin{equation}
	\label{eq:DastD}
	\begin{aligned}
		\slashed{D}^{\ast} \slashed{D} & = \frac{1}{2} \left(\left\{\gamma^\mu, \gamma^\nu\right\} + \left[\gamma^\mu, \gamma^\nu \right] \right) D_\mu D_\nu \\
		& = D^2 + \frac{1}{2} \left(e F_{\mu \nu} + \eta S_{\mu \nu} \gamma^5\right) \sigma^{\mu \nu}.
	\end{aligned}
	%\vspace{-10pt}
\end{equation}
\\Thus, we arrive at:\\
\vspace{-30pt}
\begin{strip}
	\raggedright
	%\vspace{-30pt}\\
	%\vspace{-30pt} % Center the line
	\rule{8cm}{0.5pt}%[0.5em]{8cm}{0.5pt}
	%\rule{\textwidth}{0.5pt} % Horizontal line spanning the text width with a thickness of 0.5pt
	%\vspace{1em}
	\begin{equation}
		\label{eq:trlnD2}
		\overline{\Gamma}[A,S] = -\frac{i}{2} \mathrm{Tr} \ \mathrm{Ln} \left[\frac{\left(i \partial + e A + \eta S \gamma^5\right)^2 - \frac{1}{2} \left(e F_{\mu \nu} + \eta S_{\mu \nu} \gamma^5\right) \sigma^{\mu \nu} + 2 \eta m \slashed{S} \gamma^5 - m^2}{\left(-\partial^2 - m^2\right)} \right],
		%\vspace{-5pt}
	\end{equation}
	\raggedleft
	%\vspace{1em} % Add some vertical space
	\hfill \rule{8cm}{0.5pt} % Another horizontal line
	\vspace{-10pt}
\end{strip}
Here, we have utilized Eq.~\eqref{eq:trlnequiv} and substituted Eq.~\eqref{eq:DastD}, with both numerator and denominator multiplied by $-1$ for convenience.

Next, we transition to an operator formalism, employing hat notation for operators temporarily. Let $\hat{x}^\mu$ and $\hat{p}_\mu$ be the position and momentum operators, respectively, which satisfy the commutation relations $[\hat{x}^\mu, \hat{p}_\nu] =-i \delta^\mu_\nu$, each possessing a complete eigenbasis $\hat{x}^\mu \ket{x} = x^\mu \ket{x}$ and $\hat{p}_\mu \ket{p} = p_\mu \ket{p}$ satisfying: 
\begin{equation}
	\label{eq:fourier}
	\braket{p}{x} = \frac{1}{\left(2 \pi\right)^2} e^{i p \cdot x}.
\end{equation}
Therefore, we have:
\begin{equation}
	\label{eq:heisenberg}
	\bra{x} \hat{p}_\mu \ket{\Psi} = i \partial_\mu \braket{x}{\Psi}
\end{equation}
for any state $\ket{\Psi}$. We denote:
\begin{equation}
	i \hat{D}_\mu \braket{x}{\Psi} \to \ket{\Psi} = \bra{x} \left(\hat{\Pi}_\mu + \eta \hat{S}_\mu \gamma^5\right) \ket{\Psi},
\end{equation}
where $\hat{\Pi}_\mu = \hat{p}_\mu + e \hat{A}_\mu$, and we will drop the hat notation for the remainder of this paper. By employing the identities \cite{lounestoclifford}:
\begin{equation}
	\label{eq:gamma_prod}
	\hspace{-7pt}
	\begin{aligned}
		\scalemath{0.86}{\gamma^5 \sigma^{\mu \nu} =-\frac{i}{2} \varepsilon^{\mu \nu \alpha \beta} \sigma_{\alpha \beta}} \ \ \text{and} \ \ \scalemath{0.86}{\gamma^5 \gamma^\mu =-\frac{1}{6} \varepsilon^{\mu \nu \alpha \beta} \gamma_{\nu} \sigma_{\alpha \beta}},
	\end{aligned}
	\vspace{-10pt}
\end{equation}
we simplify Eq. \eqref{eq:trlnD2} to:\\
\vspace{-10pt}
\begin{strip}
	\raggedright
	%\vspace{-10pt}\\ % Center the line
	\rule{8cm}{0.5pt}%[0.5em]{8cm}{0.5pt}
	%\rule{\textwidth}{0.5pt} % Horizontal line spanning the text width with a thickness of 0.5pt
	%\vspace{1em}
	\begin{equation}
		\label{eq:gammaAS}
		\overline{\Gamma}[A,S] =-\frac{i}{2} \mathrm{Tr} \ \mathrm{Ln} \left[\frac{\left(\Pi_\mu + \eta S_\mu \gamma^5\right)^2 - \frac{1}{2} \bigg(V_{\mu \nu} - \frac{2 \eta m}{3} \varepsilon_{\alpha \beta \mu \nu} S^\alpha \gamma^\beta \bigg) \sigma^{\mu\nu} - m^2}{p^2 - m ^2} \right],
		%\vspace{-5pt}
	\end{equation}
	\raggedleft
	%\vspace{1em} % Add some vertical space
	\hfill \rule{8cm}{0.5pt} % Another horizontal line
	%\vspace{-10pt}
\end{strip}
\vspace{-10pt}
\\where we define $V_{\mu \nu} = e F_{\mu \nu} - i \eta \tilde{S}_{\mu \nu}$, $\tilde{S}^{\mu \nu} = \frac{1}{2} \varepsilon^{\alpha \beta \mu \nu} S_{\alpha \beta}$, and it is noteworthy that any tilde over a tensor operator henceforth will be interpreted as its Hodge dual. Now, we proceed to employ Fock--Schwinger's proper-time representation \cite{schwinger1951,Fock} for every $\tau > 0$:%, provided that analytic continuation exists:
\begin{equation}
	\label{eq:propertime}
	\hspace{-4pt}
	\scalemath{1.0}{\overline{\Gamma} \left[A, S \right] = \frac{i}{2} \int_0^{\infty} \frac{d \tau}{\tau} e^{-i \tau m^2} \mathrm{Tr} \left[e^{i \tau \mathcal{H}}-e^{i \tau p^2}\right]},
\end{equation}
%\vspace{-10pt}
where $\mathcal{H}$ denotes the quadratic Hessian functional:
\begin{equation}
	\label{eq:Hessian}
	\hspace{-6pt}
	\scalemath{0.83}{\mathcal{H} = \left(\Pi_\mu + \eta S_\mu \gamma^5\right)^2 - \frac{1}{2} \bigg(V_{\mu \nu} - \frac{2 \eta m}{3} \varepsilon_{\alpha \beta \mu \nu} S^\alpha \gamma^\beta \bigg) \sigma^{\mu\nu}}.
\end{equation}
This representation brings forth a long and well discussed \cite{schwinger1951,schwartz} analogy between these exponentials in Eq. \eqref{eq:propertime} and time-evolution operators such as $e^{i \tau \mathcal{H}} \ket{x} = \ket{x; \tau}$. In this context, $\tau$ acts as a form of ``proper time,'' while the remaining terms function akin to a ``Hamiltonian'' that adheres to its own Schr\"{o}dinger-like functional equations, expressed as:
\begin{equation}
	\label{eq:schrodinger}
	-i \partial_\tau \braket{y;0}{x; \tau} = \bra{y; 0} \mathcal{H} \ket{x; \tau}.
\end{equation}
Here, $\partial_\tau = \partial/\partial \tau$.

Moreover, the correlation between the last term in Eq. \eqref{eq:propertime} and the coincidence limit of the free particle's propagator facilitates the application of standard integration techniques within momentum representation \cite{Sakurai}. This yields:
\begin{equation}
	\label{eq:propagfree}
	\begin{gathered}
		\scalemath{0.97}{\mathrm{Tr} \, e^{i \tau p^2} = \int d^4 x \bra{x} e^{i \tau p^2} \ket{x} \mathrm{tr} \, \mathbf{1} = \int d^4 x \, \frac{i}{(2 \pi \tau)^{2}}}.
	\end{gathered}
\end{equation}
In this equation, $\mathrm{tr} \, \mathbf{1} = 4$ represents the trace in Dirac space. In contrast, the first term in Eq. \eqref{eq:propertime} presents complexities due to its Hessian, which necessitates a detailed analysis that will be conducted in the following section.

\section{Diagonalization and Taxonomy} \label{sec:hessian}%Sector Classification}

In this section, we present the Hessian's diagonalization procedure and the taxonomy of its sectors. We begin by decomposing the first term of the Hessian as follows:
\begin{equation}
\label{eq:D2}
\hspace{-6pt}
\scalemath{0.98}{\left(\Pi + \eta S \gamma^5 \right)^2 = \left(\Pi^2 + \eta^2 S^2\right) \, \mathbf{1} + \eta \left\{\Pi_\mu, S^\mu \right\} \gamma^5}.
\end{equation}
Here, $S^2 = S_\mu S^\mu$. %, and it is noteworthy that any operator product will henceforth be interpreted as anti-commutators; for instance,
%\begin{equation}
%\label{eq:PiS}
%2 \Pi_\mu S^\mu \equiv \left\{\Pi_\mu, S^\mu \right\} = \Pi_\mu S^\mu + S^\mu \Pi_\mu.
%\end{equation}
We can recast Eq.~\eqref{eq:Hessian} as $\mathcal{H} = (\Pi^2 + \eta^2 S^2) \mathbf{1} + \mathcal{H}_\mathrm{R}$. The remaining term is defined as:
\begin{equation}
\label{eq:Hremainder}
\hspace{-8pt}
\scalemath{0.85}{\mathcal{H}_\mathrm{R} = \eta \left\{\Pi_\mu, S^\mu \right\} \gamma^5 - \frac{1}{2} \bigg(V_{\mu \nu} - \frac{2 \eta m}{3} \varepsilon_{\alpha \beta \mu \nu} S^\alpha \gamma^\beta \bigg) \sigma^{\mu\nu}}.
\end{equation}
This term features a characteristic equation\footnotemark[1]:
\begin{equation}
\label{eq:charac_eq}
\lambda^4 + 4 \mathcal{A} \lambda^2 + 16 i \mathcal{B} \lambda + 4 \mathcal{C} = 0.
\end{equation}
The definitions of its elements can be found in the Appendix \ref{app:solution}. The solutions take the form:

\footnotetext[1]{Obtained by means of a Python code written by the author \cite{diagmine} and solved using Wolfram's Mathematica~\cite{MathematicaNotebook,Eigenvecs}.}
\begin{equation}
\label{eq:sol_gen}
\lambda^{(\pm)}_{\pm} = \left(\pm \right) \frac{\sqrt{6}}{6} \left[\sqrt{\mathcal{D}_{-}} \pm i \sqrt{\mathcal{D}_{+} \left(\pm \right) \frac{24 i \mathcal{B}}{\sqrt{\mathcal{D}_{-}}}} \right],
\end{equation}
where the notations $(\pm)$ and $\pm$ represent independent signs, with $\mathcal{D}_{\pm}$ [Eq. \eqref{eq:Deep_into_chaos}] being complicated functions of:
\begin{equation}
\label{eq:invars1}
\begin{gathered}
	\scalemath{0.97}{\mathcal{V} = \frac{1}{4} V^2_{\mu \nu} = e^2 \mathcal{F}_{A} + \eta^2 \mathcal{F}_{S}-\frac{i e \eta}{2} \widetilde{S}^{\mu \nu} F_{\mu \nu}}, \\
	\scalemath{0.97}{\mathcal{W} =-\frac{1}{4} \widetilde{V}^{\mu \nu} V_{\mu \nu} = e^2 \mathcal{G}_{A} + \eta^2 \mathcal{G}_{S} + \frac{i e \eta}{2} S^{\mu \nu} F_{\mu \nu}},
\end{gathered}
\end{equation}
with the usual fundamental invariants of the electromagnetic field denoted as $\mathcal{F}_A = F_{\mu \nu} F^{\mu \nu}/4$ and $\mathcal{G}_A =-\tilde{F}^{\mu \nu} F_{\mu\nu}/4$ while the axial-vector analogues are specified as $\mathcal{F}_S = S_{\mu \nu} S^{\mu \nu}/4$ and $\mathcal{G}_S =-\tilde{S}^{\mu \nu} S_{\mu \nu}/4$.

From Eqs. \eqref{eq:sol_gen} and \eqref{eq:Deep_into_chaos}, certain singular points~\cite{exceptionalpoints} can exist where the Hessian ceases to be diagonalizable, exhibiting a sensitivity characteristic of non-Hermitian degeneracies~\cite{degeneracies}. Specifically, this might occur when $\mathcal{D}_{-} \to 0$ or $\infty$, indicating configurations where (i) $\Delta \to 0$ [Eq. \eqref{eq:discr}], (ii) $\Delta \to \infty$, (iii) $\mathcal{A} \to \infty$, and (iv) $\mathcal{B} \to \infty$ are extremely sensitive and possibly unstable, suggesting potential symmetry breaking or phase transitions~\cite{PhaseTransitions}. It is important to note that eigenvalue degeneracy alone does not guarantee non-diagonalizability, as some regions manifest as standard diabolical points with a complete basis of eigenvectors. However, true Exceptional Points do emerge in specific parameter subspaces. For instance, despite the computational difficulty in obtaining the eigenvector expressions for the general case, we find for $\mathcal{V} = \mathcal{W} = (\tilde{V}^{\mu \nu} S_\nu)^2 = 0$ that the system exhibits specific exceptional points. In this regime, by appealing to the quasi-static approximation established before, the anticommutator $\{\Pi_\mu, S^\mu\}$ %---itself a well-defined, self-adjoint operator on the Hilbert space, being the anticommutator of the Hermitian operators $\Pi_\mu$ and $S^\mu$---
effectively commutes with the remaining spinorial matrix structure of $\mathcal{H}_R$. This ensures that both operators can be simultaneously diagonalized to a first approximation, allowing us to treat $\{\Pi_\mu, S^\mu\}$ algebraically as a c-number within the spinorial subspace. Under these well-defined conditions, the system presents the eigenvalues:
\begin{equation}
\lambda_{\pm} = \pm \eta \sqrt{\left(\left\{\Pi_\mu, S^\mu\right\}\right)^2 - 4 m^2 S^2} \ ,
\end{equation}
with corresponding right-eigenvectors: 
\begin{equation}	
\hspace{-8pt}
\begin{aligned}
	\scalemath{0.86}{\ket{R_{1,\pm}}} & \scalemath{0.86}{= \left(\frac{2 m S_z + 2 \lambda_{\pm}}{\left\{\Pi_\mu, S^\mu\right\} - 2 m S_t}, \frac{2 m (S_x + i S_y)}{\left\{\Pi_\mu, S^\mu\right\} - 2 m S_t}, 1, 0\right)} \, , \\
	\scalemath{0.86}{\ket{R_{2,\pm}}} & \scalemath{0.86}{= \left( \frac{2 m (S_x - i S_y)}{\left\{\Pi_\mu, S^\mu\right\} - 2 m S_t}, \frac{-2 m S_z + 2 \lambda_{\pm}}{\left\{\Pi_\mu, S^\mu\right\} - 2 m S_t}, 0, 1 \right)} \ .
\end{aligned}
\end{equation}
At the threshold when the inverse $(\{\Pi_\mu, S^\mu\} - 2 m S_t)^{-1}$---now legitimately defined as an ordinary scalar quantity within this simultaneous eigenbasis, rather than as a merely formal operator expression---exhibits a pole, these eigenvectors become pairwise linearly dependent, strictly reducing the geometric multiplicity of the eigenspace. Additionally, we observe that there exist certain regions in the parameter space at which the Hessian loses Hermiticity, i.e., when $\lambda$ has non-vanishing imaginary part (e.g. $\mathcal{B} = 0$ and $\mathcal{D}_{\pm} \in \mathbb{R}$), signaling instabilities. While this may initially seem unphysical for effective actions derived from fundamental theories (such as Quantum Gravity and Extensions of the Standard Model), it could prove beneficial for deriving particle creation rates within these theories, or be applicable in other domains, such as Open Systems \cite{OpenSystems} and PT-Symmetric models \cite{Hambiorthogonal}, which are useful in Quantum Optics \cite{PTsymOptics}, Nuclear Physics \cite{MOISEYEV,ComplexHamiltonianApplication}, and beyond.

The parameter space can be systematically categorized into various types based on the limiting regions where the system displays distinct properties:

%\vspace{-10pt}
\paragraph{Extended Quantum Electrodynamics (EQED):} 
In this regime, the conditions $\left\{\Pi_\mu, S^\mu \right\} = 2 \widetilde{V}^{\mu \nu} S_\nu = 0$ and $S^2 \simeq 0$ hold, leading to the simplification of the Hessian to $\mathcal{H}_\mathrm{R} = \frac{1}{2} V_{\mu \nu} \sigma^{\mu \nu}$. This yields expressions that are analogous to those found in QED, where the quantity $V^{\mu \nu}$ serves as an extended version of the Faraday tensor $F_{\mu \nu}$. The eigenvalues associated with this system can be expressed as follows:
\begin{equation}
\label{eq:Eqed}
\lambda^{(\pm)}_{\pm} = (\pm) \sqrt{2 \mathcal{V} \pm 2 i \mathcal{W}}. 
\end{equation}

\vspace{-25pt}
\paragraph{Mixed-Transverse (MT):} In the MT case, the fields are constrained by the condition $\left\{\Pi_\mu, S^\mu \right\} =  0$. Provided that $e^{\tau \mathcal{H}}$ is trace-class and both $S_\mu$ and $A_\mu$ are smooth fields that vanish sufficiently rapidly at infinity, we find that $\left\{\Pi_\mu, S^\mu \right\} \to - i \partial_\mu S^\mu + 2 e A_\mu S^\mu = 0$ holds, indicating that either
\begin{enumerate}[label=(\alph*)]
\item The axial field is complex. In this scenario, the constraint necessitates that it possesses an electric charge of $Q = 2 e $, transforming as $S_\mu \to e^{-2 i e \alpha} S_\mu$ when $A_\mu$ transitions to $A_\mu + \partial_\mu \alpha$. This ensures that $\left\{\Pi_\mu, S^\mu \right\} = 0$ remains valid across any gauge choice. Consequently, the Proca action needs to be adjusted to the form:
\begin{equation}
	\scalemath{1.0}{\mathrm{S}_{\text{charged}} = \int d^4 x \left(- \frac{1}{2} S^{\ast}_{\mu \nu} S^{\mu \nu} + M^2_S S^2\right),} 
\end{equation}
where $S^2 = S^{\ast}_\mu S^\mu$ and $S_{\mu \nu} = D^{A}_\mu S_\nu - D^{A}_\nu S_\mu$, with the gauge covariant derivative defined as $D^{A} \equiv \partial - 2 i e A$, among other necessary modifications. A noteworthy outcome is that the covariant divergence of the Proca equation for the axial vector, sourced by a classical chiral current, now takes the form:
\begin{equation}
	M^2_S D^{A}_\mu S^\mu = \eta D^{A}_\mu j^\mu_5 + \frac{i e}{2} F_{\mu \nu} S^{\mu \nu}. 
\end{equation}

\item Alternatively, both $S_\mu$ and $A_\mu$ may be real fields. In this case, they are required to be transverse and orthogonal, specifically satisfying $A_\mu S^\mu = \partial_\mu (S^\mu) = 0$.
\end{enumerate}

%\break

Under any of these scenarios, the eigenvalues take the form:
\begin{equation}
\label{eq:MT-type}
\hspace{-6pt}
\scalemath{0.8}{\lambda^{(\pm)}_{\pm} = \left(\pm\right) \sqrt{2 \mathcal{V}-4 m^2 \eta^2 S^2 \pm 2 \sqrt{4 m^2 \eta^2 \left(\widetilde{V}^{\mu \nu} S_\nu\right)^2-\mathcal{W}^2}}}. 
\end{equation}
The system remains well-behaved as long as the condition $\mathcal{V} + \Re \sqrt{4 m^2 \eta^2 (\widetilde{V}^{\mu \nu} S_\nu)^2-\mathcal{W}^2} \geq 2 m^2 \eta^2 S^2$ is satisfied. Three special cases are noteworthy within this subspace: First, the EQED-type is encompassed within this broader subspace. Second, If $\mathcal{V} = \mathcal{W} = 0$, Eq. \eqref{eq:MT-type} implies:
\begin{equation}
\label{eq:MTchiral-asym}
\lambda^{(\pm)}_{\pm} = \left(\pm\right) 2 \sqrt{\eta m \left(-\eta m S^2 \pm \abs{\widetilde{V}^{\mu \nu} S_\nu} \right)}. 
\end{equation}
%\vspace{10pt}
When the pseudo-vector is light-like ($S^2 = 0$), the full Hessian becomes:
\begin{equation}
\label{eq:MT-type2}
\hspace{-9pt}
\scalemath{0.8}{\mathcal{H} =} \begin{bmatrix}
	\scalemath{0.77}{\Pi^2 \, \mathbf{1} + 2 \sqrt{\eta m\abs{\widetilde{V}^{\mu \nu} S_\nu}} \, \sigma_z} & \scalemath{0.77}{0} \\
	\scalemath{0.77}{0} & \scalemath{0.77}{\Pi^2 \, \mathbf{1} + 2 i \sqrt{\eta m\abs{\widetilde{V}^{\mu \nu} S_\nu}} \, \sigma_z} 
\end{bmatrix}, 
\end{equation}
where $\sigma_z = \mathrm{diag} \{1,-1\}$ represent the Pauli matrix in the $z$-direction. %\vspace{25pt}%Third, when $(\widetilde{V}^{\mu \nu} S_\nu)^2 = 0$ and the generalized fields are (anti-)self-dual, i.e., $\mathcal{V} = \pm i \mathcal{W}$, it follows that

%\vspace{-25pt}
\paragraph{Non-anomalous (NonAnom):}
The CP violating contributions are encoded by the terms $\mathcal{G}_A$, $\mathcal{G}_S$ (both inside of $\mathcal{W}$) and $\tilde{S}^{\mu \nu} F_{\mu \nu}$ (inside $\mathcal{V}$), as can be seen in the Table \eqref{tab:CP-parities}. In the region where $\mathcal{W} = 0$, we find the eigenvalues given by:\\

\vspace{-10pt}
\begin{strip}
\raggedright
\vspace{-25pt}\\ % Center the line
\rule{8cm}{0.5pt}%[0.5em]{8cm}{0.5pt}
%\rule{\textwidth}{0.5pt} % Horizontal line spanning the text width with a thickness of 0.5pt
%\vspace{1em}
\begin{equation}
	\label{eq:almost-not-anomalous}
	\lambda^{(\pm)}_{\pm} = \left(\pm\right) \sqrt{2 \mathcal{V} + \eta^2 \left[\left(\left\{\Pi_\mu, S^\mu \right\}\right)^2 - 4 m^2 S^2 \right] \pm 2 \eta \sqrt{\left(\left\{\Pi_\mu, S^\mu \right\}\right)^2 \mathcal{V} + 4 m^2 \left(\widetilde{V}^{\mu \nu} S_\nu\right)^2}}. 
	%\vspace{-5pt}
\end{equation}
\raggedleft
%\vspace{1em} % Add some vertical space
%\hfill \rule{8cm}{0.5pt} % Another horizontal line
\vspace{-10pt} % Add some vertical space
\end{strip}
%\vspace{-10pt}

The invariant $\mathcal{W}$ decomposes into a CP-even piece plus a CP-odd piece, therefore this regime can manifest under the conditions that $S^{\mu \nu} F_{\mu \nu} = \tilde{S}^{\mu \nu} F_{\mu \nu} = 0$ and $e^2 \mathcal{G}_A = - \eta^2 \mathcal{G}_S$, enabling both particles and antiparticles to exhibit consistent CP properties. Alternatively, either particles or antiparticles may present $\mathcal{W} = 0$, with the other having $\mathcal{W} \neq 0$ at the same time, provided the vector components are allowed to be complex numbers and $e^2 \mathcal{G}_A + \eta^2 \mathcal{G}_S = - i e \eta/2 S^{\mu \nu} F_{\mu \nu}$. Notably, in the first scenario, it corresponds to disconnected vector and pseudo-vector sectors, which propagate in configurations for which any local CP violation arising from one field is counterbalanced by an opposite CP violation from the other field. This dynamic results in an (almost) preserved CP symmetry; for instance, a circularly polarized light ray is associated with an ``axial-light ray'' from $S_\mu$ in the opposite polarization\footnote[2]{Further discussion of this kind of behavior can be found at sections D.3 and D.4 of Hehl and Obukhov (2012) \cite{Hehlelectro}.}. When $S_\mu$ is massive, CP violations are confined to its characteristic scale ($\sim 1/M_S$), thus allowing CP conservation to be inherited by the freely propagating $A_\mu$ at larger distances. Conversely, in cases where either particles or antiparticles are constrained to $\mathcal{W} = 0$, the resultant phenomenology indicates a more intricate CP-asymmetric selection that is contingent on the coupling between their respective field strengths, and hence particles and their CP-conjugates can exhibit different eigenvalue spectra and pair-production rates (cf. Sec. \ref{sec:Apps}).

However, there are important considerations: the UV calibration of the theory must be approached with caution to avoid deeper issues such as ghosts \cite{torsionconsistency}. Generally, the underlying model cannot be fundamental, and the mass of the axial vector must be significantly larger than that of the fermion to prevent instability in the IR sector. For a very massive axial-vector, given the equations \eqref{eq:Full_act}, \eqref{eq:Proca_act}, and \eqref{eq:Dirac_act}, along with the quasi-static assumption at low energies ($\partial_\mu S^{\mu \nu} \approx 0$), we then have:
\begin{equation}
\frac{\delta \mathrm{S} \left[A, S\right]}{\delta S_\mu} \approx S^\mu + \frac{\eta}{M^2_S} j_5^\mu = 0, 
\end{equation}
where $j_5^\mu = \langle \olsi{\psi} \gamma^\mu \gamma^5 \psi \rangle$. This leads us to conclude that:
\begin{equation}
\abs{\left\{\Pi_\mu, S^\mu \right\}} \propto \frac{m}{M^2_S} \approx 0. 
\end{equation}
From this relation, we can further simplify the eigenvalues, yielding:
\begin{equation}
\label{eq:reduced-non-anomalous}
\lambda^{(\pm)}_{\pm} \approx \left(\pm\right) \sqrt{2 \mathcal{V}-4 \eta^2 m^2 S^2 \pm 4 m \eta \abs{\widetilde{V}^{\mu \nu} S_\nu}}. 
\end{equation}
If we take into account the condition under which $|V^{\mu \nu} S_\nu|$ $= 0$, we can simplify this expression even further to:
\begin{equation}
\label{eq:reduced-non-anomalous2}
\lambda^{(\pm)}_{\pm} \approx \left(\pm\right) \left(\sqrt{2 \mathcal{V}} \pm 2 i \eta m S\right). 
\end{equation}
This reduction is justified by the relation that connects the quantities involved, specifically:
\begin{equation}
\label{eq:tVsrelation}
(\widetilde{V}^{\mu \nu} S_\nu)^2 =-2 S^2 \mathcal{V} + (V^{\mu \nu} S_\nu)^2. 
\end{equation}
These eigenvalues retain some degree of stability under the condition $\mathcal{V} + 2 \eta m |\widetilde{V}^{\mu \nu} S_\nu| \geq 2 \eta^2 m^2 S^2$ and are amenable to analysis since they avoid the complication of handling $\left\{\Pi_\mu, S^\mu \right\}$ within a square root. In addition, under the condition that $\left\{\Pi_\mu, S^\mu \right\}$ is not negligible while $|V^{\mu \nu} S_\nu| = 0$, we obtain the relation:
\begin{equation}
\label{eq:reduced-non-anomalous3}
\hspace{-6pt}
\scalemath{0.98}{\lambda^{(\pm)}_{\pm} = \left(\pm\right) \left[\eta \sqrt{\left(\left\{\Pi_\mu, S^\mu \right\}\right)^2 - 4 m^2 S^2} \pm \sqrt{2 \mathcal{V}} \right]}. 
\end{equation}
This expression can be further simplified in the case where the pseudovector takes on a light-like configuration.

%\vspace{-15pt}
\paragraph{Null Field Configurations (NFC):} 
We now turn our attention to the specific subspace where the algebraic invariants constructed from $V_{\mu\nu}$ vanish, namely $\mathcal{V} = \mathcal{W} = 0$. From the definition of these invariants, this limit enforces the strict constraints, from \eqref{eq:invars1}:
\begin{equation}
e^2 \mathcal{F}_{A} + \eta^2 \mathcal{F}_{S} = 0, \qquad \tilde{S}^{\mu \nu} F_{\mu \nu} = 0. 
\end{equation}
In classical electrodynamics, a field tensor satisfying these exact constraints is classified as a \textit{null field} (or a null electromagnetic configuration), famously representing the invariant structure of plane waves. Apart from this, Eq. \eqref{eq:almost-not-anomalous} simplifies to:
\begin{equation}
\label{eq:RMW-torsion}
\hspace{-8pt}
\scalemath{0.84}{\lambda^{(\pm)}_{\pm} = \left(\pm\right) \sqrt{\eta^2 \left(\left\{\Pi_\mu, S^\mu \right\}\right)^2 - 4 \eta^2 m^2 S^2 \pm \eta m \abs{\widetilde{V}^{\mu \nu} S_\nu}}}. 
\end{equation}
When the condition $|\widetilde{V}^{\mu \nu} S_\nu| = 0$ is satisfied, the Hessian of a system lying within this subspace can be hermitian if and only if $2 m S \leq |\left\{\Pi_\mu, S^\mu \right\}|$. A special simplification arises for a pseudo-vector in light-like configurations ($S^2 = 0$), which can be achieved either (i) classically if it is massless or (ii) off-shell if it is quantized. In such cases, the Hessian takes the form:
\begin{equation}
\label{eq:RMW-H}
\mathcal{H} = \begin{bmatrix}
	\left(\Pi_\mu + \eta S_\mu\right)^2 \mathbf{1} & 0 \\
	0 & \left(\Pi_\mu-\eta S_\mu\right)^2 \mathbf{1}
\end{bmatrix}, 
\end{equation}
where $\mathbf{1}$ denotes a 2-by-2 identity matrix block. If $|\widetilde{V}^{\mu \nu} S_\nu| \neq 0$ but $M_S \gg m$, we find again Eq. \eqref{eq:MT-type2}.

In time-like configurations of the axial vector, the Hessian exhibits these same signs of instability of Eq. \eqref{eq:MTchiral-asym}, even for a very massive axial vector. This implies that the eigenvalues invariably contain at least two pure imaginary components, provided the condition $|\widetilde{V}^{\mu \nu} S_\nu| \geq \eta m S^2$ holds. %If attainable, these scenarios indicate a \emph{chirality-asymmetric stability}, where parity is violated in a manner that allows one chirality sector to remain ``safe'' while the opposite sector invariably decays into particles\footnote{This safety can be put into question if higher loops, interactions, and backreactions are included according to more complicated scenarios.}. This interplay results in a \emph{chirality-asymmetric pair creation}, characterized by a net production of fermions with a preferred chirality. Should any process propel $\eta m S^2$ beyond the remaining term, the system will become entirely unstable, precipitating rapid vacuum decay.

Significantly, this sector and the similar region within MT, where Eq. \eqref{eq:MT-type2} is valid, also demonstrates a kinematical symmetry reduction to the $SIM(2)$---or $ISO(2)$ if the norm of $S_\mu$ is fixed---subgroup of the Lorentz group. This assertion is underscored by the structure of Eqs. \eqref{eq:MT-type2} and \eqref{eq:RMW-H}, which bears resemblance to that found in Very Special Relativity \cite{VSRGlashow}. Similar symmetry breaking is also observed in the preceding subspaces, though not with the same level of explicitness and specificity. Consequently, models constructed within this subspace have the potential to exhibit a rich phenomenology, including the intriguing possibility of imparting mass to neutrinos without necessitating a Seesaw mechanism \cite{VSRmass}, provided that careful considerations are made during the model's development\footnotemark[3].

%\vspace{-5pt}
\footnotetext[3]{While the Hessian in question is well-behaved, it may not encompass all desired representations, including projective ones, as discussed by da Rocha et. al. \cite{Marcondes,Marcondes2}.}

\paragraph{Asymmetric Self-Dual (ASD):} This region is characterized by either of the conditions (a) $\{\Pi_\mu, S^\mu\} = 0$ or (b) $S = 0$, both of which are accompanied by (anti-)self-dual fields: %renomear para Asymmetric (anti-)Self-Dual (ASD) se os editores não gostarem
\begin{equation}
\tilde{V}_{\mu \nu} = \pm i V_{\mu \nu} \implies e F_{\mu \nu} = \pm \eta S_{\mu \nu} = g W_{\mu \nu}, 
\end{equation}
where $W_{\mu \nu} = \partial_\mu W_\nu - \partial_\nu W_\mu$. In particular, we can express this expression as:
\begin{equation}
V_{\mu \nu} = \frac{g}{2} (g_{\alpha \mu} g_{\beta \nu} - g_{\alpha \nu} g_{\beta \mu} \pm i \varepsilon_{\mu \nu \alpha \beta}) W^{\alpha \beta}, 
\end{equation}
with $g_{\mu \nu}$ representing the metric tensor. Consequently, by decomposing the axial-vector into its transverse and longitudinal components, denoted as $S_\mu = S^{\perp}_\mu + S^{\parallel}_\mu$, it follows that $e A_\mu = \pm \eta S^{\perp}_\mu = g W_\mu$. Thus, within this subspace, the covariant derivative can be articulated as:
\begin{equation}
D_\mu = \partial_\mu - i g W_\mu \mathrm{P}_{H} - i \eta S^{\parallel}_\mu \gamma^5, 
\end{equation}
where $H = R$ ($L$) if the fields exhibit (anti-)self-duality. The decision to use the notation $g W_\mu$ was made intentionally to highlight the structural similarity to the projector configuration associated with the Standard Model's $W$-boson post symmetry breaking.

In this sector, Eq.~\eqref{eq:tVsrelation} leads us to the expression:
\begin{equation}
\left(V^{\mu \nu} S_\nu \right)^2 = S^2 \mathcal{V} = - \left(\tilde{V}^{\mu \nu} S_\nu \right)^2. 
\end{equation}
From this, we can derive the eigenvalues of the Hessian and their corresponding stability conditions:
\begin{enumerate}[label=(\alph*)]
\item For the overlap region with the MT subspace, we obtain:
\begin{equation}
	\label{eq:MT-selfdual}
	\hspace{-12pt}
	\scalemath{0.82}{\lambda^{(\pm)}_{\pm} = (\pm) \sqrt{2 \left( \mathcal{V} \mp \sqrt{ \mathcal{V}^2  + 4 m^2 \eta^2 S^2  \mathcal{V}} \right) - 4 m^2 \eta^2 S^2}}, 
\end{equation}
with stability conditions expressed as:
\begin{equation}
	\left( \mathcal{V}  \mp \sqrt{ \mathcal{V}^2  + 4 m^2 \eta^2 S^2  \mathcal{V}}\right) \geq 2 m^2 \eta^2 S^2. 
\end{equation}
This indicates that the vacuum for one chirality is stable when: 
\begin{equation}
	\label{eq:stabilityASD1}
	\begin{cases}
		- \mathcal{V} \leq 4 m^2 \eta^2 S^2 \leq 8 \mathcal{V}  & \text{when } \mathcal{V} > 0, \\
		S^2 \leq 0, & \text{when } \mathcal{V}  \leq  0, 
	\end{cases} 
\end{equation}
while the other chirality remains stable for: 
\begin{equation}
	\label{eq:stabilityASD2}
	\begin{cases}
		- \mathcal{V} \leq 4 m^2 \eta^2 S^2 \leq 0 & \text{when } \mathcal{V} > 0, \\
		m^2 \eta^2 S^2 \leq 2 \mathcal{V}, & \text{when } \mathcal{V} \leq 0. 
	\end{cases} 
\end{equation}
Notably, two interesting scenarios can arise: (i) For $\mathcal{V} > 0$ and $S^2 > 0$, the vacuum of one chirality is always \textit{unstable}, whereas the other features a mixture of stable and unstable regions; and (ii) when $\mathcal{V} \leq 0$ and $S^2 \leq 0$, the situation is inverted---one chirality is perpetually \textit{stable} while the other can exhibit both stable and unstable zones.

\item Additionally, we observe that, for $S^2 = 0$:
\begin{equation}
	\label{eq:Aselfdual}
	\hspace{-18pt}
	\begin{aligned}
		\scalemath{0.9}{\mathcal{H} =} & \begin{bmatrix}
			\scalemath{0.9}{\left(\Pi_\mu + \eta S_\mu\right)^2 \mathbf{1}} & 0 \\
			0 & \scalemath{0.9}{\left(\Pi_\mu - \eta S_\mu\right)^2 \mathbf{1} + 2 \sqrt{\mathcal{V}} \sigma_z}
		\end{bmatrix}, \\
		& = \begin{bmatrix}
			\scalemath{0.9}{p^2 \mathbf{1}} & 0 \\
			0 & \scalemath{0.9}{\left(p_\mu + g W_\mu\right)^2 \mathbf{1} + 2 \sqrt{\mathcal{V}} \sigma_z}
		\end{bmatrix}
	\end{aligned} 
\end{equation}
which ensures complete stability for the vacuum of one chirality. However, the other chirality can become unstable when $\mathcal{V} < 0$.
\end{enumerate}
Both regions converge when $\left\{\Pi_\mu, S^\mu\right\} = S = 0$, leading to a simplified version of the Hessian given by:
\begin{equation}
\label{eq:Aselfdualintersec}
\mathcal{H} = \begin{bmatrix}
	p^2 \mathbf{1} & 0 \\
	0 & \Pi^2_{(W)} \mathbf{1} + 2 \sqrt{\mathcal{V}} \sigma_z
\end{bmatrix}, 
\end{equation}
where, notably, $\Pi_{(W)} \equiv p + g W$.

Crucially, it is evident that Eqs. \eqref{eq:MTchiral-asym}, \eqref{eq:MT-type2}, \eqref{eq:RMW-H}, \eqref{eq:MT-selfdual}, and \eqref{eq:Aselfdual} exhibit chiral-asymmetric properties, presenting the possibility of intriguing realizations of Very Special Relativity (VSR).
%\vspace{30pt}

%\break

\paragraph{Other regions:} Apart from the previously discussed subspaces, further regions of interest may exist elsewhere in parameter space, contingent upon careful considerations and specific choices informed by the structure of Eq. \eqref{eq:sol_gen}. Given that the combined regions of MT-and NonAnom-types encompass a significant portion of the parameter space (even intersecting each other), the remaining regions are either special cases within these two categories---such as those involving the vanishing of additional invariants besides those whose vanishing defines the previously listed subspaces (e.g., the subspace within NonAnom where $S$ also vanishes) or potentially exhibiting intricate eigenvalue structures. There are also noteworthy regions that exhibit long eigenvalue expressions. Those are strictly governed by conditions such as $S = 0$, $\mathcal{V} = 0$, $(V^{\mu \nu} S_\nu)^2 = 0$, or any combinations thereof, without imposing additional constraints on other invariants.\\
\vspace{-30pt}

%\hspace{-20pt}
\begin{strip}
\centering
%\vspace{-10pt}\\ % Center the line
%\rule{8cm}{0.5pt}%[0.5em]{8cm}{0.5pt}
%\begin{table}%[h]
\captionof{table}{Summary of the sector taxonomy presented here.\label{tab:subspaces}}
\vspace{5pt}
\setlength{\tabcolsep}{2pt}
\renewcommand{\arraystretch}{3.5}
\resizebox{0.95\textwidth}{!}{\hspace{-30pt}%
	\begin{tabular}{|lccc|}
		\hline
		Type & $\left(\lambda^{(\pm)}_{\pm}\right)^2$ & Stability & Theory Constraint \\ \hline
		EQED & $2 \mathcal{V} \pm 2 i \mathcal{W}$ & $\sim$ QED's conditions & \makecell[c]{Massless $S_\mu$ \\ or like-wise configurations} \\ \arrayrulecolor{black}
		\hline
		MT & $2 \mathcal{V}-4 m^2 \eta^2 S^2 \pm 2 \sqrt{4 m^2 \eta^2 \left(\widetilde{V}^{\mu \nu} S_\nu\right)^2-\mathcal{W}}$ & \makecell[c]{$2 m^2 \eta^2 S^2 \leq \mathcal{V} +$ \\ $+ \Re \sqrt{4 m^2 \eta^2 \left(\widetilde{V}^{\mu \nu} S_\nu\right)^2-\mathcal{W}^2}$ \\ Possibly Chirality-Asymmetric} & All theories \\
		\hline
		NonAnom & \makecell[c]{$2 \mathcal{V} + \eta^2 \left[\left(\left\{\Pi_\mu, S^\mu \right\}\right)^2 - 4 m^2 S^2\right] \pm$ \\
			$\pm 2 \eta \sqrt{\left(\left\{\Pi_\mu, S^\mu \right\}\right)^2 \mathcal{V} + 4 m^2 \left(\widetilde{V}^{\mu \nu} S_\nu\right)}$} & $m \ll M_S$ plus same as MT & Massive $S_\mu$ \\
		\hline
		NFC & $\eta^2 \left(\left\{\Pi_\mu, S^\mu \right\}\right)^2 - 4 \bigg[\eta^2 m^2 S^2 \mp \eta m \abs{\widetilde{V}^{\mu \nu} S_\nu}\bigg]$ & \makecell[c]{$\eta \left(\left\{\Pi_\mu, S^\mu \right\}\right)^2 + 4 m \abs{\widetilde{V}^{\mu \nu} S_\nu} \geq 4 \eta m^2 S^2$ \\ Possibly Chirality-Asymmetric} & \makecell[c]{All theories \\ (Better for massless $S_\mu$)} \\
		\hline
		\multirow{2}{*}{ASD \makecell[c]{(a) \\ \\ \\ \\ (b)}} & \makecell[c]{$2 \left(\mathcal{V} \mp \sqrt{\mathcal{V}^2 + 4 m^2 \eta^2 S^2 \mathcal{V}}\right) - 4 m^2 \eta^2 S^2$\\ $ \ $ } & \makecell[c]{$\mathcal{V} \mp \sqrt{\mathcal{V}^2 + 4 m^2 \eta^2 S^2 \mathcal{V}} \geq 2 m^2 \eta^2 S^2$\\
			Possibly Chirality-Asymmetric} & All theories, \\
		& \makecell[c]{$\left(\sqrt{2 (1 \pm 1) \mathcal{V}} \mp 2 \eta \left\{\Pi_\mu, S^\mu \right\}  \right)^2$ \\ $ \ $} & \makecell[c]{$\mathcal{V} > 0$ \\ Possibly Chirality-Asymmetric} & Massless $S_\mu$ \\
		\hline
		Others & $\frac{1}{6} \left[\sqrt{\mathcal{D}_{-}} \pm i \sqrt{\mathcal{D}_{+} \left(\pm \right) \frac{24 i \mathcal{B}}{\sqrt{\mathcal{D}_{-}}}} \right]^2$ & Needs careful handling & All theories \\ \arrayrulecolor{black} \hline
	\end{tabular}%
}
%\vspace{5pt}
%\captionof{table}{Summary of the sector taxonomy presented here.}
%\label{tab:sectors}
%\end{table}
%\raggedleft
%\vspace{-25pt}\\ % Center the line
%\rule{8cm}{0.5pt}%[0.5em]{8cm}{0.5pt}
\end{strip}
%\vspace{-10pt}

A summary of the relevant subspaces can be found in Table \ref{tab:subspaces}. We conclude this section by noting that the MT-and ASD-type represents the consistent subspaces for both massive and massless axial vectors. For MT-type subspace, it is required that the inequality
\begin{equation}
\scalemath{0.96}{\mathcal{V} + \Re \sqrt{4 m^2 \eta^2 \left(\widetilde{V}^{\mu \nu} S_\nu\right)^2-\mathcal{W}^2} \geq 2 \eta^2 m^2 \eta m S^2} 
\end{equation}
holds true. Importantly, this subspace reduces to the EQED-type for massless axial-vector theories where $S^2 = 0$ and share common regions with both NonAnom-and NFC-type subspaces. The latter remains fully stable for Electroaxial theories involving massless pseudo-vectors, presenting intriguing and physically consistent characteristics for other massive axial-vector theories. Nonetheless, NFC-, ASD-and MT-type permits the potential for chirality-asymmetric pair creation when a system with massive $S_\mu$ evolves into it from another subspace. Conversely, massive pseudo-vector theories within the NonAnom-type (including NFC-type) may not be considered fundamental---primarily due to the presence of $\left\{\Pi_\mu, S^\mu \right\}$ within $\mathcal{H}_\mathrm{R}$---and the condition $M_S \gg m$ must be satisfied along with constraints on $\mathcal{V}$ and/or $|\widetilde{V}^{\mu \nu} S_\nu|$ to ensure the existence of at least one stable chirality sector.

Further stability constraints may be required for different scenarios where $\mathrm{S}_0 [A, S]$ could affect the consistency requirements or in cases where spacetime 
%\vspace{-1.5pt}
curvature is taken into account, or when both vector fields are quantized~\cite{StabilityOfVector,Nonpertrenorm,StabilitydeRham}.

\section{The One-Loop Effective Actions\label{sec:exact}}

In this section, we derive closed-form expressions for effective actions of models situated within the previously classified consistent subspaces by employing zeta-function regularization. This process involves taking the trace in the eigenbasis of the Hessian and subsequently applying minimal subtraction (MS).

We will initiate our analysis by by observing that from Eq.~\eqref{eq:reduced-non-anomalous3} and Eq.~\eqref{eq:Aselfdual}, it is evident that within the framework of ASD-b and the specific subspaces of NonAnom and NFC [as expressed in Eq.~\eqref{eq:RMW-H}], where $S = |\widetilde{V}^{\mu \nu} S_\nu| = 0$, the Hessian can be represented as follows:
\begin{equation}
\label{eq:1stclassH0}
\hspace{-8pt}
\scalemath{0.84}{\mathcal{H} =} \begin{bmatrix}
	\scalemath{0.84}{\left(\Pi_\mu + \eta S_\mu\right)^2 \mathbf{1} + \mathcal{X}_{+} \sigma_z} & 0 \\
	0 & \scalemath{0.84}{\left(\Pi_\mu - \eta S_\mu\right)^2 \mathbf{1} + \mathcal{X}_{-} \sigma_z }
\end{bmatrix}, 
\end{equation}
with $\mathcal{X}^{\rm NFC}_{\pm} = \mathcal{X}^{\rm ASD}_{+} = 0$ and/or $\sqrt{2} \mathcal{X}^{\rm NonAnom}_{\pm} = \mathcal{X}^{\rm ASD}_{-} = 2 \sqrt{\mathcal{V}}$. Furthermore, the remaining previously reported subspaces can be described more generally as follows:
\begin{equation}
\label{eq:2stclassH}
\mathcal{H} = \left(\Pi^2 + \eta^2 S^2 \right) \mathbf{1} + X, 
\end{equation}
where $X = \mathrm{diag} \{\mathcal{X}_{+} \sigma_z, \mathcal{X}_{-} \sigma_z\}$ and $\mathcal{X}_{\pm} = |\lambda^{\left(\pm\right)}_{\pm}|$. This distinction allows us to categorize the previously determined Hessians into two distinct classes, which we shall systematically identify regarding their eigenbasis on a case-by-case basis. We will begin with the simplest case, described by Eq.~\eqref{eq:1stclassH0}, which we will denote as \textit{First Class}. The more complex case represented by Eq.~\eqref{eq:2stclassH} will subsequently be addressed and classified as \textit{Second Class}.

%\vspace{-15pt}	
\subsection{First class:}

In evaluating the effective action for this sector, where both fields are massless (or in a likewise configuration), we must define the background configuration for the geometric field $S_\mu$. Unlike the electromagnetic field, $S_\mu$ does not possess gauge symmetry; thus, a gauge-fixing procedure is inapplicable. Instead, we take advantage of the quasi-static approximation. By Taylor expanding the field around the origin, $S_\mu (x) \approx S_\mu (0) + x^\alpha \partial_\alpha S_\mu(0) + \dots$, and considering this specific configuration where the field is null ($S^2 = 0$), we can set the origin such that $S_\mu(0) = 0$. Furthermore, in this quasi-static regime, we assume the symmetric part of the derivative is negligible or strictly constant, leaving the anti-symmetric contribution as the dominant term. This allows us to parameterize the background as $S_\mu \approx -\frac{1}{2} S_{\mu\nu} x^\nu$. While this parameterization mathematically mimics the structure of the Fock-Schwinger gauge, it is physically justified here as a linearized background approximation for null, quasi-static fields, facilitating the algebraic decoupling in the trace computation. As for the electromagnetic field, we can utilize Fock-Schwinger's gauge directly. The chiral potentials then become $V^{(R/L)}_\mu =-$ $(e F_{\mu \nu} \pm \eta S_{\mu \nu}) x^\nu/2$ $\pm$ $\bar{S}_\mu$. If $F_{\mu \nu}$ and $S_{\mu \nu}$ form a regular pencil\footnote[4]{See, e.g., the Gantmacher (1974) \cite{gantmacher} for further details.}, it is possible to simultaneously transform them into Jordan's form through an appropriate choice of coordinates. By applying boosts and rotations, we can choose a reference frame where each vector potential can be parameterized as:
\begin{equation}
V^{(H)}_\mu = \left(-E^{(H)}_{V} x, 0,-B^{(H)}_{V} z, 0\right), 
\end{equation}
where $(H) = (R)$ or $(L)$ (with $\mathcal{X}_{(R/L)} \equiv \mathcal{X}_{\pm}$) indicates the chirality sector. Here, the quantities $E^{(H)}_V$ and $B^{(H)}_V$ represent the chiral electric and magnetic field analogs extracted from $V^{(H)}_{\mu \nu} = \partial_\mu V^{(H)}_{\nu}-\partial_\nu V^{(H)}_{\mu}$, for which the relevant invariants are closely related to Eq. \eqref{eq:invars1}:
\begin{equation}
\label{eq:invars2}
\hspace{-4pt}
\begin{gathered}
	\scalemath{0.93}{\mathcal{F}^{(H)}_V = \frac{1}{4} V^{(H)}_{\mu \nu} V_{(H)}^{\mu \nu} = e^2 \mathcal{F}_{A} + \eta^2 \mathcal{F}_{S} \pm \frac{e \eta}{2} S^{\mu \nu} F_{\mu \nu}}, \\
	\scalemath{0.93}{\mathcal{G}^{(H)}_V =-\frac{1}{4} \widetilde{V}^{(H)}_{\mu \nu} V_{(H)}^{\mu \nu} = e^2 \mathcal{G}_{A} + \eta^2 \mathcal{G}_{S} \pm \frac{e \eta}{2} \widetilde{F}^{\mu \nu} S_{\mu \nu}},
\end{gathered} 
\end{equation}
and therefore:
\begin{equation}
\label{eq:electromagV}
\hspace{-3pt}
\begin{aligned}
	\scalemath{0.82}{\left(B^{(H)}_{V}\right)^2-\left(E^{(H)}_{V} \right)^2 = 2 \mathcal{F}^{(H)}_V} \ \ \text{and} \ \ \scalemath{0.82}{\vec{E}^{(H)}_{V} \cdot \vec{B}^{(H)}_{V} = \mathcal{G}_V^{(H)}}.
\end{aligned} 
\end{equation}
Furthermore, in the case of the special subspace inside NFC, since $\mathcal{V} = \mathcal{W} = 0$, whereby we find $E^{(H)}_V = \pm B^{(H)}_V$ when $\mathcal{F}^{(H)}_V = \mathcal{G}^{(H)}_V = 0$, that is, they are in the class of null field configurations, regardless of their specific spatio-temporal profile. Alternatively, in ASD subspace, if the fields are allowed to take on complex values, we find $B^{(H)}_V = i E^{(H)}_V$.% = \sqrt{\mathcal{F}^{(H)}_V}$.

Thus, in general, we can infer that Eq. \eqref{eq:2stclassH} reduces to: 
\begin{equation}
\label{eq:1stclassH}
\hspace{-6pt}
\scalemath{0.83}{\mathcal{H}_{\mathrm{H}} = \left[p_t-E^{(H)}_{V} x \right]^2-p^2_x-\left[p_y-B^{(H)}_{V} z \right]^2-p_z ^2 + \mathcal{X}_{(H)} \sigma_z.} 
\end{equation}
Next, by decomposing the $n$-th eigenstate of $\mathcal{H}$ into its chirality components $\ket{\psi_n} = \ket{\psi_{R,n}} + \ket{\psi_{L,n}}$, we can express the trace as $\mathrm{Tr} e^{-i \tau \mathcal{H}} = \mathrm{Tr}_{\mathrm{R}} e^{-i \tau \mathcal{H}_{\mathrm{R}}} + \mathrm{Tr}_{\mathrm{L}} e^{-i \tau \mathcal{H}_{\mathrm{L}}}$ and employ momentum operators as follows:
\begin{equation}
\left[p_t-E^{(H)}_{V} x \right]^2 = e^{\frac{i p_t p_x}{E^{(H)}_{V}}} \left[E^{(H)}_{V} x\right]^2 e^{-\frac{i p_t p_x}{E^{(H)}_{V}}}, 
\end{equation}
without the need to modify $\mathcal{X}_{(H)}$ where the quasi-static approximation holds, allowing us to further reduce Eq. \eqref{eq:2stclassH} into quantum harmonic oscillators, such that:
\begin{equation}
\label{eq:1stclassH2}
\hspace{-6pt}
\scalemath{0.8}{\mathcal{H}_{\mathrm{H}} =-\left[p^2_x-\left(E^{(H)}_{V} \right)^2 x^2\right]-\left[p^2_z + \left(B^{(H)}_{V} \right)^2 z^2\right] + \mathcal{X}_{(H)} \sigma_z.} 
\end{equation}

\vspace{-10pt}

\subsection{Second class:}

For the massive case, the background parameterization requires a careful physical distinction. Any vector field can be decomposed into its transverse and longitudinal components, $S_\mu = S_\mu^\perp + S_\mu^\parallel$, such that $S^2 = (S^\perp)^2 + (S^\parallel)^2$. In the regime where the quasi-static assumption holds and the longitudinal modes dominate the dynamics, i.e., $S^2 \approx (S^\parallel)^2 = \bar{S}^2_\mu$, the transverse fluctuations $(S^\perp)^2$ can be treated as a negligible perturbation. Under this assumption, we have the freedom to parameterize the transverse configuration without altering the leading-order physics. We then separate the field as $S_\mu \approx -\frac{1}{2} S_{\mu\nu} x^\nu + \bar{S}_\mu$. This is a choice of background parameterization when the longitudinal mode dominates, instead of an imposition of a gauge. We can then define the effective mass: 
\begin{equation}
\label{eq:eff_mass}
M^2 = m^2-\eta^2 \bar{S}^2 
\end{equation}
which reconfigures Eq. \eqref{eq:2stclassH} into:
\begin{equation}
\label{eq:2stclassH2}
\mathcal{H} = \left(p_\mu-\frac{1}{2} e F_{\mu \nu} x^\nu \right)^2 + \frac{1}{4} \eta^2 \left(S_{\mu \nu} x^\nu\right)^2 + X. 
\end{equation}
This expression can be simplified under the conditions that either the matrix representations of both strength tensors commute and their Jordan forms exhibit non-zero blocks at the same positions, or that each tensor contains only one non-vanishing block. Conversely, if we parameterize the fields and apply the momentum operator, we obtain:
\begin{equation}
\label{eq:proof}
\hspace{-7pt}
\scalemath{0.87}{e^{-i \frac{p_t p_x}{a}}\left[\left(p_t-a x\right)^2 + b^2 t^2\right]e^{i \frac{p_t p_x}{a}} = a^2 x^2 + b^2 \left(t + \frac{p_x}{a}\right)^2}, 
\end{equation}
which essentially reinstitutes the original form, rather than yielding the desired result. Hence, we assume that the strength tensors possess appropriate Jordan blocks and define the analogous electric and magnetic fields $E_S$ and $B_S$ derived from the transverse axial vector. We subsequently denote the genuine electric and magnetic fields $E_A$ and $B_A$ derived from the vector potential $A_\mu$, such that the potentials can be parameterized as $A_\mu = (-E_A x, 0,-$ $B_A z, 0)$ and $S_{\mu}^{\perp} = (-E_S x, 0,-B_S z, 0)$ or interchangeably shift $z \to x$ for either vector. It follows that both the genuine and analogue fields adhere to relations similar to those in Eq. \eqref{eq:electromagV}, specifically:  
\begin{equation}
\label{eq:electromagX}
\begin{aligned}
	\scalemath{0.94}{B^2_{A/S}-E^2_{A/S} = 2 \mathcal{F}_{A/S} \quad} \text{and} \quad \scalemath{0.94}{\vec{E}_{A/S} \cdot \vec{B}_{A/S} = \mathcal{G}_{A/S}}.
\end{aligned} 
\end{equation}
%\vspace{7.5pt}
We can then apply the momentum operators as to shift: 
\begin{equation}
\label{eq:shift}
\begin{aligned}
	x \to x + \frac{e E_A p_t}{E^2} & & \text{ and } \quad z \to z + \frac{e B_A p_y}{B^2},
\end{aligned} 
\end{equation}
%\vspace{7.5pt}
where we define the composite electromagnetic fields:
\begin{equation}
\label{eq:EletromagNoIndex}
\hspace{-6pt}
\begin{aligned}
	E^2 = e^2 E^2_{A} + \eta^2 E^2_{S} & & \text{ and } \quad B^2 = e^2 B^2_{A} + \eta^2 B^2_{S}.
\end{aligned} 
\end{equation}
This transformation renders Eq. \eqref{eq:2stclassH2} into the following form:
\begin{equation}
\label{eq:2stclassH3}
\hspace{-6pt}
\scalemath{0.88}{\mathcal{H} = \frac{\eta^2 E^2_S}{E^2} p^2_t-\frac{\eta^2 B^2_S}{B^2} p^2_y-p^2_x + E^2 x^2-p^2_z-B^2 z^2 + X}, 
\end{equation}
%\vspace{7.5pt}
which is reminiscent of a Hamiltonian featuring two seemingly ``massive'' free directions, each characterized by different ``mass'' parameters, alongside the oscillatory components.

\vspace{7.5pt}
\subsection{The Generalized Euler-Heisenberg Actions:}

Now we can address both Hessian classes in a unified framework. Each class will exhibit Landau Levels \cite{Landau} as eigenstates, with energy eigenvalues:
\begin{equation}
\label{eq:energies}
\hspace{-6pt}
\scalemath{0.91}{\mathcal{E}^{(H), \lambda, c}_{n_{\scalemath{0.6}{E}}, n_{\scalemath{0.6}{B}}} = i (1 + 2 n_{\scalemath{0.6}{E}}) E^{(H)}_{I} + (1 + 2 n_{\scalemath{0.6}{B}}) B^{(H)}_{I} + 2 \lambda \mathcal{X}_{(H)}}. 
\end{equation}
%\vspace{7.5pt}
In this context, $\lambda = \pm \frac{1}{2}$ denotes the spin eigenvalues, while the index $I$ serves as a dummy variable that varies according to the specific physical framework considered. This index can represent different types of subsystems, taking the forms of $A$, $S$, $V$, or can be omitted entirely. These variations are employed to characterize the genuine, axial analogue, chiral [Eq.~\eqref{eq:electromagV}], and composite [Eq.~\eqref{eq:EletromagNoIndex}] electromagnetic fields, respectively. At this stage, we proceed to execute a standard textbook calculation\footnotemark[5].

%\vspace{7.5pt}
To facilitate our analysis, we assume integration within a Euclidean box of size $L$ and calculate the number of eigenstates contained within: (i) for the first class, each chirality sector may possess a different density of eigenstates, approximately given by $\sim 4 \pi^2/(\mathcal{G}^{(H)}_V L^2)$, whereas (ii) for the second class, at energies below the fermion mass, the factors multiplying the momenta in Eq. \eqref{eq:2stclassH3} will have a minimal impact on the dispersion relation, leading to a density of eigenstates approximately $\sim 4 \pi^2/(\mathcal{G} L^2)$. Consequently, we derive:%, where $\mathcal{G} = E B$.
%\pagebreak
\vspace{-10pt}
\begin{strip}
\raggedright
%\vspace{-15pt}\\ % Center the line
\rule{8cm}{0.5pt}\\%[0.5em]{8cm}{0.5pt}
\vspace{15pt}
%\rule{\textwidth}{0.5pt} % Horizontal line spanning the text width with a thickness of 0.5pt
%\vspace{1em}
\begin{equation}
	\hspace{-10pt}
	\begin{aligned}
		\mathrm{Tr} e^{i \tau \mathcal{H}} = \sum_{(H) = (R/L)}\frac{i L^4 \mathcal{G}^{(H)}_{I}}{\left(2 \pi\right)^2} \left(\sum_{\lambda = \pm \frac{1}{2}} e^{2 i \tau \lambda \mathcal{X}_{(H)}}\right) \left[\sum_{n_{\scalemath{0.6}{E}} = 0} ^{\infty} e^{-\tau E^{(H)}_I \left(1 + 2 n_{\scalemath{0.6}{E}}\right)}\right] \left[\sum_{n_{\scalemath{0.6}{B}} = 0} ^{\infty} e^{i \tau B^{(H)}_I \left(1 + 2 n_{\scalemath{0.6}{B}}\right)}\right],
	\end{aligned} 
\end{equation}
%	\raggedleft
%\vspace{1em} % Add some vertical space
%	\rule{8cm}{0.5pt} % Another horizontal line
%	%\vspace{-10pt} % Add some vertical space
%\end{strip}
%\vspace{-10pt}
which, upon returning to the expression $L^4 \to \int d^4 x$ and Wick rotating $\tau \to-i \tau$, ultimately yields:
%\begin{strip}
%	\raggedright
%	%\vspace{-25pt}\\ % Center the line
%	\rule{8cm}{0.5pt}%[0.5em]{8cm}{0.5pt}
%\rule{\textwidth}{0.5pt} % Horizontal line spanning the text width with a thickness of 0.5pt
%\vspace{1em}
\begin{equation}
	\label{eq:genEHalmost}
	\hspace{-8pt}
	%\vspace{-10pt}
	\overline{\Gamma}_{c} \left[A, S\right] = \frac{1}{16 \pi^2} \sum_{(H)} \int d^4 x \ \frac{d \tau}{\tau} \left[\frac{ e^{-M_{c}^2 \tau} \mathcal{G}^{(H)}_{I} \cosh{\left(\tau \mathcal{X}_{H}\right)}}{\sin{\left(\tau E^{(H)}_I \right)} \sinh{\left(\tau B^{(H)}_I\right)}}-\frac{e^{-m^2 \tau}}{\tau^2}\right]. 
\end{equation}
\raggedleft\\
%\vspace{0pt}\\ % Add some vertical space
\vspace{-10pt}
\hfill \rule{8cm}{0.5pt} % Another horizontal line
\vspace{-10pt} % Add some vertical space
\end{strip}
%\vspace{10pt}
Here, we denote the mass term of class $c$ as $M_c$, specifically $M_{1} = m$ and $M_{2}$ is given by Eq.~\eqref{eq:eff_mass}. %\vspace{-20pt}%.

More explicitly, in the first class, the whole quantum correction vanishes for plane wave solutions, while in ASD-b, for instance, it takes the form:
\vspace{-30pt}
\begin{strip}
\raggedright
\vspace{-10pt} % Center the line
\rule{8cm}{0.5pt}%[0.5em]{8cm}{0.5pt}
%\rule{\textwidth}{0.5pt} % Horizontal line spanning the text width with a thickness of 0.5pt
\vspace{10pt}
\begin{equation}
	\label{eq:genEHalmost1}
	%\vspace{-10pt}
	\begin{aligned}
		\scalemath{1.0}{\overline{\Gamma}_{1} \left[A, S\right] = \frac{1}{16 \pi^2} \int d^4 x \ \frac{d \tau}{\tau} \ e^{-m^2 \tau} \left[\frac{\left(E^{(L)}_{V}\right)^2 \cosh{\left(2 \tau \sqrt{\mathcal{V}} \right)}}{\sin^2{\left(\tau E^{(L)}_{V} \right)}} - \frac{1}{\tau^2}\right]},
	\end{aligned} 
\end{equation}
\raggedleft
\vspace{-10pt}\\ % Add some vertical space
\hfill \rule{8cm}{0.5pt}
\vspace{-15pt} % Add some vertical space
\end{strip}
\vspace{0pt}
%\vspace{5pt}
\\since $V^{(R)} = 0 \implies E^{(R)}_V \to 0$, and results in:\\
%\vspace{-10pt}
%\vfill
%\break
\begin{equation}
\vspace{25pt}
\label{eq:genEHalmost2}
\hspace{-10pt}
\begin{gathered}
	\scalemath{0.82}{\overline{\Gamma}_{2} \left[A, S\right] = \frac{1}{8 \pi^2} \int d^4 x \ \frac{d \tau}{\tau} \left[e^{-M^2 \tau} \mathcal{G}\frac{\Re\cosh{\left(\tau \mathcal{X}_{+}\right)}}{\Im \cosh{\left(\tau \bar{\mathcal{X}}_{+} \right)}}-\frac{e^{-m^2 \tau}}{\tau^2} \right]}
\end{gathered} 
\end{equation}
%\footnotetext[5]{As can be seen in section 33.4 from Schwartz (2013) \cite{schwartz}.}
for the second class. For this, we have defined $\bar{\mathcal{X}}_{\pm} = \sqrt{2 \mathcal{F} \pm 2 i \mathcal{G}} = B \pm i E$, where $\mathcal{F} = (B^2-E^2)/2$. To employ a minimal subtraction scheme, we initially expand both components around $\tau \to 0$, resulting in:
\vfill

\footnotetext[5]{As can be seen in section 33.4 from Schwartz (2013) \cite{schwartz}.}

\break

%\footnotetext[5]{As can be seen in section 33.4 from Schwartz (2013) \cite{schwartz}.}
%\vspace{-30pt}
\begin{strip}
\raggedright
\vspace{-45pt}\\ % Center the line
%\rule{8cm}{0.5pt}%[0.5em]{8cm}{0.5pt}
%\rule{\textwidth}{0.5pt} % Horizontal line spanning the text width with a thickness of 0.5pt
%\vspace{1em}
\begin{equation}
	\hspace{-10pt}
	\begin{aligned}
		\scalemath{1.0}{\frac{\left(E^{(H)}_I\right)^2 \cosh{\left(\tau \mathcal{X}_{H}\right)}}{\sin^2{\left(\tau E^{(H)}_{V} \right)}} = \frac{1}{\tau^2} + \frac{1}{6} \left[2 \left(E^{(H)}_V\right)^2 + 3 \mathcal{X}_{H}^2 \right] + \frac{\tau^2}{120} \left[8 \left(E^{(H)}_V \right)^4 + 20 \left(E^{(H)}_V \right)^2 \mathcal{X}_{H}^2 + 5 \mathcal{X}_{H}^4 \right] + \dots}
	\end{aligned} 
\end{equation}
%\raggedleft
%\vspace{1em} % Add some vertical space
%\hfill\rule{8cm}{0.5pt} % Another horizontal line
%\vspace{-10pt} % Add some vertical space
%\end{strip}
\\
for each chirality sector of the first class, and:\\
%\break
\vspace{10pt}
%\begin{strip}
%\raggedright
%	%\vspace{-25pt}\\ % Center the line
%\rule{8cm}{0.5pt}%[0.5em]{8cm}{0.5pt}
%\rule{\textwidth}{0.5pt} % Horizontal line spanning the text width with a thickness of 0.5pt
%\vspace{1em}
\begin{equation}
	\begin{aligned}
		\scalemath{0.83}{\mathcal{G}\frac{\Re\cosh{\left(\tau \mathcal{X}_{+}\right)}}{\Im \cosh{\left(\tau \bar{\mathcal{X}}_{+} \right)}} = \frac{1}{\tau^2} + \frac{1}{12} \left[3 \left(\mathcal{X}^2_{+} + \mathcal{X}^2_{-}\right)-4 \mathcal{F}\right] + \frac{1}{720} \left[56 \mathcal{F}^2-8 \mathcal{G}^2-60 \mathcal{F} \left(\mathcal{X}^2_{+} + \mathcal{X}^2_{-}\right) + 15 \left(\mathcal{X}^2_{+} + \mathcal{X}^2_{-}\right)^2-30 \mathcal{X}^2_{+} \mathcal{X}^2_{-} \right] \tau^2 + \cdots}
	\end{aligned} 
\end{equation}
\raggedleft
%\vspace{1em} % Add some vertical space
\hfill \rule{8cm}{0.5pt}\\
\vspace{-10pt} % Add some vertical space
\end{strip}
\vspace{-10pt}
for the second. By subtracting the first terms, we arrive at the Generalized Euler--Heisenberg Lagrangians for equations \eqref{eq:genEHalmost1} and \eqref{eq:genEHalmost2}, which can be expressed as follows:\\
\vspace{-50pt}
\begin{strip}
\vspace{-10pt}
\raggedright
%\vspace{-10pt} % Center the line
\rule{8cm}{0.5pt}%[0.5em]{8cm}{0.5pt}
%\rule{\textwidth}{0.5pt} % Horizontal line spanning the text width with a thickness of 0.5pt
%\vspace{1em}
\begin{equation}
	\label{eq:genEH1}
	%\hspace{-10pt}
	\scalemath{1.0}{\mathcal{L}_{1} \left[A, S\right] = \frac{1}{16 \pi^2} \int \frac{d \tau}{\tau} \ e^{-m^2 \tau} \left\{\frac{\left(E^{(L)}_{V}\right)^2 \cosh{\left(2 \tau \sqrt{\mathcal{V}} \right)}}{\sin^2{\left(\tau E^{(L)}_{V} \right)}} - \frac{1}{\tau^2}-\frac{1}{3} \left[\left(E^{(L)}_V\right)^2 + 6 \mathcal{V}\right]\right\}}, 
\end{equation}
\vspace{-10pt}
%\\%\raggedleft
%\vspace{5pt} % Add some vertical space
%\hfill \rule{8cm}{0.5pt}
%\vspace{-10pt} % Add some vertical space
%\end{strip}\\
%\begin{strip}
%\raggedright
%\vspace{-20pt}\\ % Center the line
%\rule{8cm}{0.5pt}%[0.5em]{8cm}{0.5pt}
%\rule{\textwidth}{0.5pt} % Horizontal line spanning the text width with a thickness of 0.5pt
%\vspace{1em}
\begin{equation}
	\label{eq:genEH2}
	\scalemath{1.0}{\mathcal{L}_{2} \left[A, S\right] = \frac{1}{8 \pi^2} \int \frac{d \tau}{\tau} e^{-M^2 \tau} \left[ \mathcal{G}\frac{\Re\cosh{\left(\tau \mathcal{X}_{+}\right)}}{\Im \cosh{\left(\tau \bar{\mathcal{X}}_{+} \right)}}-\frac{1}{\tau^2}-\frac{1}{12} \left(3 \mathcal{X}^2_{+} + 3 \mathcal{X}^2_{-}-4 \mathcal{F}\right) \right]}. 
\end{equation}
\raggedleft\\
\vspace{-0pt}
%\vspace{5pt} % Add some vertical space
\hfill \rule{8cm}{0.5pt}\\ % Add some vertical space
\vspace{-20pt}
\end{strip}
%\vspace{-10pt}
\\ \indent
%\vfill
Before elaborating on this result, it is important to recognize that due to $S_\mu$ being a pseudo-vector, the mixed term in $\mathcal{F}_V^{(H)}$ ($\mathcal{G}_V^{(H)}$) does (not) violate CP invariance, contrasting with the corresponding term found in $\mathcal{V}$ ($\mathcal{W}$). Thus, additional constraints, such as $S^{\mu \nu} F_{\mu \nu} = \tilde{S}^{\mu \nu} F_{\mu \nu} = 0$, may be necessary for model consistency \cite{impossibleuai}. If this constraint holds, we deduce that $\mathcal{F}^{(R)}_V = \mathcal{F}^{(L)}_V = \mathcal{V}$ and $\mathcal{G}^{(R)}_V = \mathcal{G}^{(L)}_V = \mathcal{W}$, which leads to the conclusion that Eq.~\eqref{eq:genEH1} vanishes when it occurs for NFC subspace ($\mathcal{V} = \mathcal{W} = 0$), implying there are no surviving one-loop quantum corrections to the classical action. In other words, an effective action resembling the EH form for the NFC-type subspace is exclusively feasible within CP-violating theories. If CP violation is permitted, then Eq. \eqref{eq:MT-type2} can be simplified to a form akin to Eq. \eqref{eq:genEH1}, yielding $E_V^{(R)} = E_V^{(L)} = E$. However, it is advisable to maintain the contributions from each sector distinct to accurately capture the chirality-asymmetric instability inherent in the particle creation process\footnotemark[6].
\enlargethispage{\baselineskip}
\footnotetext[6]{See Section~\ref{sec:Apps} for more details.}

%\vspace{-30pt}
For the remainder of this section, we operate within a MT-type subspace with real fields to elucidate these results, assuming $F_{\mu \nu} S^{\mu \nu} = \tilde{S}^{\mu \nu} F_{\mu \nu} = (\widetilde{V}^{\mu \nu} S_\nu)^2 = 0$ for simplicity. %, where $\mathcal{X}_{+} = (2 \mathcal{V}-4 m^2 \eta^2 S^2 + i \mathcal{W})^{1/2}$.
Thus, Eq. \eqref{eq:genEH2} becomes:%transforms into
%\vspace{-10pt}
%\vspace{-10pt}
%\vspace{50pt}\\%\vfill
\begin{equation}
%\vspace{-10pt}
\label{eq:genEHMT}
%\vspace{30pt}
\begin{gathered}
	\scalemath{0.86}{\mathcal{L}_{MT} \left[A, S\right] = \frac{1}{8 \pi^2} \int \frac{d \tau}{\tau} e^{-M^2 \tau} \left[ \mathcal{G}\frac{\Re\cosh{\left(\tau \mathcal{X}_{+}\right)}}{\Im \cosh{\left(\tau \bar{\mathcal{X}}_{+} \right)}} - \right.} \\ \scalemath{0.86}{\left. - \frac{1}{\tau^2}-\frac{2}{3} \left(e^2 \mathcal{F}_A + \eta^2 \mathcal{F}_S-2 \eta^2 m^2 S^2 \right) \right]}, 
\end{gathered}
\end{equation}\\
and the fourth-order term of the expansion becomes:
\begin{equation}
\label{eq:I4}
%\hspace{-5pt}
\begin{gathered}
	\mathrm{I}_4 = -\frac{4 \tau^2}{45} \left[\left(e^2 \mathcal{F}_A + \eta^2 \mathcal{F}_S \right)^2 + \right. \\ \left. + \frac{7}{4} \left(e^2 \mathcal{G}_A + \eta^2 \mathcal{G}_S\right)^2\right] + \frac{2 \tau^2}{3} \eta^4 m^4 \bar{S}^{4}_{\mu}, 
\end{gathered}
\end{equation}
which aligns with the established EH Lagrangian for QED \cite{EHoriginal,schwinger1951,DunneEH} in the limit as $\eta \to 0$.

%\vspace{-30pt}

\subsection{Perturbative Analysis}

Additionally, we can verify this through perturbative analysis \cite{Dobadolivro}, expanding Eq. \eqref{eq:eff_action} to obtain:
\vspace{-10pt}
%\\
%\vspace{-10pt}
\begin{equation}
\label{eq:pertub}
\hspace{-4pt}
\begin{aligned}
	\scalemath{0.94}{\overline{\Gamma} \left[A, S \right] = i \sum_{k=1}^{\infty} \frac{1}{k} \mathrm{Tr} \left[\left(i \slashed{\partial}-m \right)^{-1} \left(\slashed{V}_{R} \mathrm{P}_R + \slashed{V}_{L} \mathrm{P}_L\right)\right]^{k}}. 
\end{aligned} 
\vspace{-10pt}
\end{equation}
%\\
%\vspace{10pt}
Thus, by naming each $k$-th term in the sum as $\overline{\Gamma}^{(k)} \left[A, S \right]$ and maintaining careful consideration of the role of $\gamma^5$ by following the prescription by t\'{}hooft and Veltman~\cite{thooftveltman,Collins} during the application of dimensional regularization, the second-order term will be given by:\\
\break
\begin{strip}
\vspace{30pt}
%\raggedright\\
%\vspace{-10pt} % Center the line
%\rule{8cm}{0.5pt}\\%[0.5em]{8cm}{0.5pt}
%\vspace{-30pt}
%\rule{\textwidth}{0.5pt} % Horizontal line spanning the text width with a thickness of 0.5pt
%\vspace{1em}
\begin{equation}
	\label{eq:perturb2}
	\hspace{-8pt}
	\begin{aligned}
		\scalemath{0.9}{\overline{\Gamma}^{(2)} \left[A, S \right] = \frac{1}{\left(4 \pi\right)^2}\int d^4 x d^4 y \frac{d^4 p}{\left(2 \pi\right)^4} e^{i p.\left(x-y\right)} \left[ I_{1} \left(p^2 g_{\mu \nu}-p_\mu p_\nu\right) V^\mu_{R} (y) V^\nu_{R} (x) + m^2 I_{2} \left( V^\mu_{R} (y) V^L_{\mu} (x)-V^\mu_{R} (y) V^{R} _{\mu} (x)\right) \right] + \mathrm{r.t.}}, %\begin{pmatrix}
		%\scalemath{0.93}{V_{R} \leftrightarrow V_{L}} \\
		%\scalemath{0.93}{\mathrm{P}_{R} \leftrightarrow \mathrm{P}_{L}}
		%\end{pmatrix}
		%\vspace{-10pt}
	\end{aligned} 
\end{equation}
\raggedleft
\vspace{-20pt}\\ % Add some vertical space
\hfill \rule{8cm}{0.5pt} % Another horizontal line
\vspace{-10pt} % Add some vertical space
\end{strip}
%\vspace{-10pt}
where r.t. stands for the remaining terms for which $\mu \leftrightarrow \nu$ and $k_1 \leftrightarrow k_2$, while:
%\vspace{-15pt}
\begin{equation}
%\vspace{0pt}
\label{eq:formfactor}
\begin{aligned}
	{\rm I}_{1} & = \frac{\Delta}{6}-\int_0 ^1 du (1-u) u \ln{\left[1-u(1-u) \frac{p^2}{m^2}\right]}, \\ {\rm I}_{2} & = \Delta-\int_0 ^1 du \ln{\left[1-u(1-u) \frac{p^2}{m^2}\right]}
\end{aligned} 
\end{equation}
\\are the form factors and $\Delta = 2/\varepsilon + \ln{(4 \pi)} -\upgamma-\ln{(m^2/\mu^2)} \to \int_0^{\infty} d\tau e^{-\tau m^2}/\tau$, with $\upgamma$ representing Euler--Mascheroni's constant, and $\mu$ signifying the energy scale factor from dimensional regularization. The result is consequently:\\
\vspace{-20pt}
\begin{strip}
\raggedright
%\vspace{-20pt}\\ % Center the line
\rule{8cm}{0.5pt}%[0.5em]{8cm}{0.5pt}
%\rule{\textwidth}{0.5pt} % Horizontal line spanning the text width with a thickness of 0.5pt
%\vspace{1em}
\begin{equation}
	\label{eq:perturbfinal}
	\hspace{-5pt}
	\begin{aligned}
		\scalemath{0.79}{\overline{\Gamma}^{(2)} \left[A, S \right] = \frac{-1}{8 \pi^2} \int d^4 x \left[\frac{2 \Delta}{3} \left(e^2 \mathcal{F}_A + \eta^2 \mathcal{F}_S + 2 \eta^2 m^2 S^2 \right)-\eta^2 S_\mu \Box S^\mu-\frac{1}{15 m^2} \left(e^2 F_{\mu \nu} \Box F^{\mu \nu} + \eta^2 S_{\mu \nu} \Box S^{\mu \nu} + 5 \eta^2 S_{\mu} \Box^2 S^{\mu}\right) + \mathcal{O} \left(\frac{1}{m^4}\right) \right]}.
	\end{aligned} 
	%\vspace{-10pt}
\end{equation}
\raggedleft
%\vspace{-0pt}\\ % Add some vertical space
\hfill \rule{8cm}{0.5pt} % Another horizontal line
%\vspace{-10pt} % Add some vertical space
\end{strip}
\vspace{-10pt}

Thus, we conclude that Eq. \eqref{eq:perturbfinal} is consistent with the minimal subtraction term of Eq. \eqref{eq:genEHMT} in the limit of constant fields since $\Delta_M = \int d\tau \, e^{-M^2 \tau}/\tau = \Delta-\ln{(1-\eta^2 S^2/m^2)} \approx \Delta$. It is noteworthy that the third-order term also conforms to expectations, despite a naive assumption that it might not be present if the analysis is kept merely at the expansion of Eq. \eqref{eq:genEHMT}, as will be elaborated in the following section.

%\vfill

\section{From the Axial Anomaly to Vacuum Back-Reaction: The Effective Divergence of the Chiral Current\label{sec:Anom}}
%\raggedbottom
%

In this section, we transition from the strictly topological anchor of the chiral anomaly toward a broader description of the chiral response. We explore how the vacuum, polarized by the presence of both vector and axial backgrounds, undergoes a dynamical back-reaction that modifies the current divergence. This framework allows us to address the symmetry-breaking process as a dynamical phenomenon where the vacuum's self-consistency plays a central role. Should the axial field be interpreted as a gauge field associated with chiral symmetry, our results demonstrate that, beyond the anomalous breaking, a dynamical component arises from the vacuum's polarization, effectively sculpting the chiral landscape of the theory. This derivation will lay the foundation for applying the pair creation rates discussed in the subsequent section to a simplified model of baryogenesis.

\subsection{Perturbative Consistency and the Anomaly Core}

Before venturing into the non-perturbative regime of the exact effective action, it is essential to check for consistency once more. This subsection is dedicated to the evaluation of the axial anomaly via triangle diagrams to ensure that our subsequent functional derivations remain grounded in the well-established requirements of chiral symmetry breaking. This calculation serves to calibrate our tools, verifying that the fundamental topological core of the anomaly is correctly recovered before we introduce the complexities of vacuum back-reaction and non-quasi-static dynamics.

\begin{figure}[h!] % Placement specifiers: here, top, bottom, page of floats
\centering
%\vspace{-20pt}
\includegraphics[width=1.0\columnwidth]{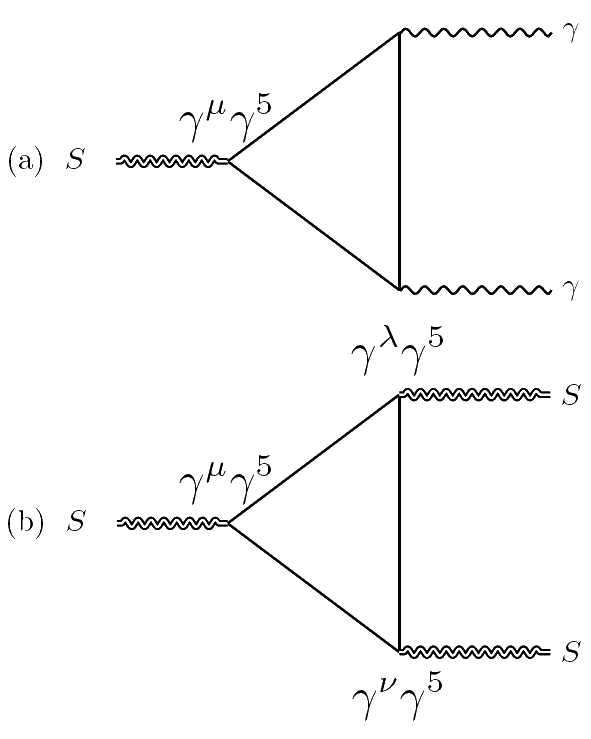}
\vspace{5pt}
\caption{\justifying
	Anomalous triangle diagrams; generated with TikZ-Feynman~\cite{TikZFeynman}; of processes (a) $\scalemath{0.85}{S \scalemath{0.85}{\to} \gamma \gamma}$ ($+$ permutations) and (b) $\scalemath{0.85}{S \scalemath{0.85}{\to} SS}$.}
\label{fig:triangle}
\end{figure}

Following Eq. \eqref{eq:pertub}, we observe that at the third order, two terms survive while the others cancel due to Furry's theorem. The former correspond to the anomalous diagrams depicted in Fig. \ref{fig:triangle}. Specifically, they represent the linearly divergent integrals:
\vspace{-20pt}
\begin{strip}
\raggedright
%\vspace{-20pt} % Center the line
\rule{8cm}{0.5pt}%[0.5em]{8cm}{0.5pt}
%\rule{\textwidth}{0.5pt} % Horizontal line spanning the text width with a thickness of 0.5pt
%\vspace{-20pt}
\begin{equation}
	\label{eq:perturb3rd}
	\hspace{-2pt}
	\begin{aligned}
		\scalemath{0.95}{\overline{\Gamma}^{(3)} \left[A, S\right] = \frac{-i \eta}{3} \int d^4 x \, d^4 y \, d^4 z \frac{d^4 k_1}{\left(2 \pi\right)^4} \frac{d^4 k_2}{\left(2 \pi\right)^4} \frac{d^4 p}{\left(2 \pi\right)^4} e^{i k_1 \cdot x + i k_2 \cdot y} \delta \left(p-k_1-k_2\right) \left[3 e^2 A^\mu_x A^\nu_y S^\tau_z I_{\mu \nu \tau} + \eta^2 S^\mu_x S^\nu_y S^\tau_z I^{(5)}_{\mu \nu \tau}\right]},
	\end{aligned} 
\end{equation}
\raggedleft
%\vspace{-10pt}\\ % Add some vertical space
%\hfill \rule{8cm}{0.5pt}\\ % Another horizontal line
%\vspace{-10pt} % Add some vertical space
%\justifying
\end{strip}
%\begin{strip}
\noindent
where we adopt:\\
\begin{strip}
%\raggedright
\vspace{-20pt}\\ % Center the line
%\rule{8cm}{0.5pt}%[0.5em]{8cm}{0.5pt}
%\rule{\textwidth}{0.5pt} % Horizontal line spanning the text width with a thickness of 0.5pt
%\vspace{1em}
\begin{equation}
	\hspace{-18pt}
	\begin{gathered}
		\scalemath{0.96}{I_{\mu \nu \tau} \left(k_1, k_2 \right) = \int \frac{d^D \tilde{q}}{\left(2 \pi\right)^D} \ \mathrm{tr} \left[\frac{i}{\slashed{q}-m} \gamma_\tau \gamma_5 \frac{i}{\slashed{q}-\slashed{p}-m} \gamma_\nu \frac{i}{\slashed{q}-\slashed{k}_1-m} \gamma_\mu \right] +} \begin{bmatrix}
			\scalemath{0.96}{\mu \leftrightarrow \nu} \\
			\scalemath{0.96}{k_1 \leftrightarrow k_2}
		\end{bmatrix} \scalemath{0.96}{= I_{(1)\mu \nu \tau} \left(k_1\right) + I_{(2)\mu \nu \tau} \left(k_2\right)},
	\end{gathered} 
\end{equation}
%\raggedleft
%\vspace{-10pt}\\ % Add some vertical space
%\hfill \rule{8cm}{0.5pt} % Another horizontal line
%\vspace{30pt} % Add some vertical space
%\end{strip}
%\vspace{30pt} % Add some vertical space
%\break
%\begin{strip}
%	\raggedright
%\vspace{25pt}\\ % Center the line
%\rule{8cm}{0.5pt}%[0.5em]{8cm}{0.5pt}
%\rule{\textwidth}{0.5pt} % Horizontal line spanning the text width with a thickness of 0.5pt
%\vspace{1em}
\begin{equation}
	\begin{gathered}
		\scalemath{0.93}{I^{(5)}_{\mu \nu \tau} \left(k_1, k_2\right) = \int \frac{d^D \tilde{q}}{\left(2 \pi\right)^D} \mathrm{tr} \left[\frac{i}{\slashed{q}-m} \gamma_\tau \gamma_5 \frac{i}{\slashed{q}-\slashed{p}-m} \gamma_\nu \gamma_5 \frac{i}{\slashed{q}-\slashed{k}_1-m} \gamma_\mu \gamma_5 \right] +} \begin{bmatrix}
			\scalemath{0.93}{\mu \leftrightarrow \nu} \\
			\scalemath{0.93}{k_1 \leftrightarrow k_2}
		\end{bmatrix} \scalemath{0.93}{= I^{(5)}_{(1) \mu \nu \tau} \left(k_1\right) + I^{(5)}_{(2) \mu \nu \tau} \left(k_2\right)}.
	\end{gathered} 
\end{equation}
\raggedleft
%\vspace{-10pt}\\ % Add some vertical space
\hfill \rule{8cm}{0.5pt} % Another horizontal line
%\vspace{-10pt} % Add some vertical space
\end{strip}
%\vspace{-20pt}\\
%\vfill \break
\vspace{-10pt}

%\begin{figure}[h!]
%	\centering
%\begin{wrapfigure}[20]{c}[-15pt]{1.0\textwidth}
%\vspace{10pt}
%	\includegraphics[scale=0.8]{Fig1.pdf}
%\captionsetup{hangindent=0pt, singlelinecheck=false, margin={0cm,-1.5cm}}
%	\caption{\justifying
%		Anomalous triangle diagrams corresponding to the (a) $\scalemath{0.9}{S \scalemath{0.9}{\to} \gamma \gamma}$ ($+$ permutations) and (b) $\scalemath{0.9}{S \scalemath{0.9}{\to} SS}$ processes.}
%	\label{fig:triangle2}
%\end{figure}
%\end{wrapfigure}

%\vspace{-15pt}

%\vspace{-5pt}
%\noindent
By applying a Taylor series expansion and performing separated translations for each integration block $I_{(\cdot), \mu \nu \tau}$ from the first diagram, one can derive\footnotemark[7]:
\begin{equation}
I_{\mu \nu \tau} \left(a, b\right)-I_{\mu \nu \tau} \left(0, 0\right) =-\frac{i \varepsilon_{\lambda \tau \nu \mu} \left(a-b\right)^\lambda}{8 \pi^2}. 
\end{equation}
\footnotetext[7]{This procedure is explored in more detail at the chapter 4 of Dobado, Maroto and Pelaez (1997) \cite{Dobadolivro}.}
Thereby, by imposing Ward's identities for vector current conservation, i.e., $\partial_\mu j^\mu = 0 \implies k_1^\mu I_{\mu \nu \tau} = k^\mu_2 I_{\mu \nu \tau} = 0$, in conjunction with the classical axial current continuity equation for massive fermions, given by $\partial_\mu j^\mu_5 = 2 i m j_5 \implies p^\lambda I_{\mu \nu \lambda} = 2 i m I_{\mu \nu}-\Delta_{\mu \nu}$, where:
\begin{equation}
\hspace{-9pt}
\begin{aligned}
	\scalemath{0.89}{I_{\mu \nu} = \scalemath{0.88}{\int \frac{d^D \tilde{q}}{\left(2 \pi\right)^D} \mathrm{tr} \left[\frac{i}{\slashed{q}-m} \gamma_5 \frac{i}{\slashed{q}-\slashed{p}-m} \gamma_\nu \frac{i}{\slashed{q}-\slashed{k}_1-m} \gamma_\mu \right] + \mathrm{r.t.}}},
	%\begin{bmatrix}
	%\mu \leftrightarrow \nu \\
	%k_1 \leftrightarrow k_2
	%\end{bmatrix}
\end{aligned} 
\end{equation}
which is finite, and:
\begin{equation}
\hspace{-9pt}
\begin{aligned}
	\scalemath{0.79}{\Delta_{\mu \nu}} & \scalemath{0.77}{ = \int \frac{d^D \tilde{q}}{\left(2 \pi\right)^D} k^\alpha \frac{\partial}{\partial q^\alpha} \mathrm{tr} \left[\frac{1}{\slashed{q}-\slashed{k}_2-m} \gamma_5 \gamma_\nu \frac{1}{\slashed{q}-\slashed{k}_1-m} \gamma_\mu \right] + \mathrm{r.t.,}} \\ %\begin{bmatrix}
	%\mu \leftrightarrow \nu \\
	%k_1 \leftrightarrow k_2
	%\end{bmatrix}
	& \scalemath{0.9}{= \frac{1}{2 \pi^2} \varepsilon_{\alpha \nu \beta \mu} k^\alpha_2 k^\beta_1},
\end{aligned} 
\end{equation}
we can derive the well-known Adler--Bell--Jackiw anomaly \cite{adler,belljackiw} (or Carroll--Field--Jackiw-Torsion \cite{paganellyTorsion} if the axial vector derives from torsion):
\begin{equation}
\label{eq:diaga}
\partial_\mu \left(j_5 ^\mu\right)_{(a)} = 2 i m \left(j_5\right)_{(a)}-\frac{e^2}{8 \pi^2} \tilde{F}^{\mu \nu} F_{\mu \nu}. 
\end{equation}
A similar procedure can be applied to the other surviving diagram; however, in this case, one must impose symmetry between vertices instead of vector-current conservation to obtain:
\begin{equation}
\label{eq:diagb}
\partial_\mu \left(j_5 ^\mu\right)_{(b)} = 2 i m \left(j_5\right)_{(b)}-\frac{\eta^2}{24 \pi^2} \tilde{S}^{\mu \nu} S_{\mu \nu}. 	
\end{equation}
This indicates that the anomaly cancellation conditions of the Standard Model are sufficient to guarantee cancellation in this context as well \cite{anomcondit,anomalycancelation}. In perturbation theory the chiral Ward identity obtained from symmetric routing of triangular amplitudes (or equivalently from a variation of the effective action computed with a non-covariant regulator) yields the consistent form of the anomaly~\cite{anomaliesinQFT,Fujikawa}, which corresponds to the expressions in Eqs. \eqref{eq:diaga} and \eqref{eq:diagb}, which induces the Wess-Zumino~\cite{wesszumino} term that needs to be added to the effective action, namely:%\\
%\vspace{-25pt}

\break

\begin{strip}
\raggedright
\vspace{-25pt}\\ % Center the line
%\rule{8cm}{0.5pt}%[0.5em]{8cm}{0.5pt}
%\rule{\textwidth}{0.5pt} % Horizontal line spanning the text width with a thickness of 0.5pt
%\vspace{1em}
\begin{equation}
	\label{eq:anominduced}
	\overline{\Gamma}^{(3)} \left[A, S \right] =-\frac{\eta}{4 \pi^2} \int d^4 x \left[e^2 S_\mu A_\nu \tilde{F}^{\mu \nu}-\frac{\eta^2}{6} \int d^4 y \left(\partial_{\alpha} S^{\alpha}\right)_x {\Box}^{-1}_{x y} \left(S_{\mu \nu} \tilde{S}^{\mu \nu} \right)_y \right], 
\end{equation}
\raggedleft
\vspace{-10pt}\\ % Add some vertical space
\hfill \rule{8cm}{0.5pt} % Another horizontal line
\vspace{-10pt}\\ % Add some vertical space
\end{strip}
%\vspace{-10pt}
\noindent
where we adopt a mixed representation featuring a non-local component, expressed as $\Box^{-1}_{x y} = (2\pi)^{-4} \int d^4 p \, e^{-i p \cdot (x-y)}/p^2$, along with a 4D Chern-Simons-like term to underscore the topological nature of the component containing the electromagnetic field, which is a gauge field.

Furthermore, this insight presents an intriguing analogy to the pion within the weak field approximation by determining decay rates for each process (e.g., for coherent splitting):
\begin{equation}
\label{eq:decaypref}
\begin{aligned}
	\scalemath{0.93}{\Gamma \left(S \to \gamma \gamma\right) \propto \frac{e^4 m^3}{256 \pi^3}} \ \ \text{and} \ \ \scalemath{0.93}{\Gamma \left(S \to S S\right) \propto \frac{\eta^4 m^3}{2304 \pi^3}}.
\end{aligned}
\end{equation}
It can be observed that $\Gamma (S \to \gamma \gamma)/\Gamma (S \to S S) = 9 e^4/\eta^4$. Thus, if $\eta < \sqrt{3} e$, there is a preferential decay of the axial field into a pair of photons. Such dynamics may suggest a mechanism for Primordial Magnetogenesis \cite{2Bornot2B,magnetobouncing} in cosmological models derived from theories incorporating propagating torsion \cite{PGGtorsion,torsioninflation}.

A noteworthy consequence of this anomaly is as follows: Let us define $j^\mu = \langle\olsi{\psi} \gamma^\mu \psi \rangle = \langle \olsi{\psi} \gamma^\mu ({\rm P}_{R} + {\rm P}_{L}) \psi \rangle = j^\mu_R + j^\mu_L$ and $j^\mu_5 = j^\mu_R-j^\mu_L$. Since $A_\mu$ is a gauge field, its associated anomaly is inherently tied to topology, specifically, it is proportional to the winding number, which is an integer:
\begin{equation}
\label{eq:winding}
\int d^4 x \partial_\mu \left(j^\mu_5\right)_{(a)} = N_R - N_L = N_5.
\end{equation}
In contrast, here we treat $S_\mu$ as not necessarily a gauge field itself, so Eq. \eqref{eq:winding} for the $S \to SS$ decay process can assume any real value. However, the existence of a $S \to \gamma\gamma$ decay channel implies both fields share topological content through their simultaneous coupling to the fermionic field. Specifically, when the couplings satisfy the condition for preferred $S \to \gamma\gamma$ decay, the axial vector is \emph{constrained} to configurations that respect the topological structure inherited from the electromagnetic field. While $S_\mu$ need not exhibit classical instanton solutions, its fluctuations must be compatible with the winding number structure of $A_\mu$, effectively forcing the transferred particle-number asymmetry to obey the gauge field's topological quantization. This constraint is stronger in NonAnom-type subspaces, where $e^2 \mathcal{G}_A = \eta^2 \mathcal{G}_S$ when $S^{\mu \nu} F_{\mu \nu} = 0$.

\subsection{Exact Chiral Current Divergence in Quasi-Static Backgrounds}

Having established the perturbative anomaly core, we now turn to the exact effective action framework to compute the full macroscopic divergence of the chiral current in this subsection. Unlike the perturbative expansion, the Schwinger proper-time method intrinsically captures the non-perturbative vacuum polarization induced by the background fields, which manifests as a dynamical non-conservation of chirality beyond the anomaly's purely topological UV residue, providing a comprehensive description of the chiral current dynamics in slowly varying macroscopic environments.

The total effective chiral current induced in the vacuum as a response to the presence of a pseudo-vector in well behaved subspaces ($\{\Pi_\mu, S^\mu\} \approx 0$) is given by:
\begin{equation}
\avg{j_5^\mu}_{\text{eff}} = \frac{1}{\eta} \frac{\delta \overline{\Gamma}_{\text{EH}}}{\delta S_\mu}.
\end{equation}
This current is covariant by design, as the Heat-Kernel method ensures the preservation of Gauge covariance. For the sake of simplicity, let us denote the effective action as follows: 
\begin{equation}
\label{eq:EHf}
\hspace{-5pt}
\scalemath{0.9}{\overline{\Gamma} \left[A, S\right] = -\frac{i}{2} \int_0 ^\infty \frac{d \tau}{\tau} e^{- m^2 \tau} \ \mathrm{Tr} \ e^{\tau \mathcal{H}} - \int d^4 x \frac{\Delta_M}{48 \pi^2} R},
\end{equation}
where $R = (e^2 F^2_{\mu \nu} + \eta^2 S^2_{\mu \nu} - 8 \eta^2 m^2 S^2)$ is the minimal subtraction term for an MT-type subspace with real fields, which serves as our illustrative example. It is important to note that the effective action depends on the vector background solely through the field strength $F_{\mu \nu}$, which in turn ensures that the vector current remains exactly conserved, $\partial_\mu \avg{j^\mu}_{\text{eff}} = 0$, preserving the $U(1)$ gauge symmetry non-perturbatively throughout our derivation. Notably, we employ:
\begin{equation}
\hspace{-6pt}
\begin{aligned}
	\scalemath{1.0}{\mathcal{H} - m^2} & \scalemath{1.0}{= \left(i\slashed{D}^{\ast} + m\right) \left(i\slashed{D} - m\right)}.
\end{aligned} 
\end{equation}
For the first term, leveraging the cyclicity property of the functional trace yields:
\begin{equation}
\delta \left(\mathrm{Tr} \, e^{\tau \mathcal{H}}\right) = \tau \mathrm{Tr} \left(e^{\tau \mathcal{H}} \delta \mathcal{H}\right), 
\end{equation}
where:
\begin{equation}
\label{eq:deltaH}
\hspace{-4pt}
\begin{aligned}
	\scalemath{0.86}{\delta \mathcal{H}} & \scalemath{0.86}{= \eta \left[\left(i \slashed{D}^{\ast} + m\right) \gamma^\mu \gamma^5 - \gamma^\mu \gamma^5 \left(i \slashed{D} -  m\right) \right] \left(\delta S_\mu\right)}.
\end{aligned} 
\end{equation}
Using the relation $\lim_{\varepsilon \to 0} \int_0^{\infty} d\tau \ e^{i \tau (r + i \varepsilon)} = \frac{i}{r}$, we arrive at:
\begin{equation}
\hspace{-6pt}
\begin{gathered}
	\frac{1}{\eta} \frac{\delta \overline{\Gamma}_{\text{EH}}}{\delta S_\mu} = \frac{1}{2} \mathrm{tr} \left(\bra{x}\frac{1}{i\slashed{D} - m} \gamma^\mu \gamma^5 \ket{x} + \right. \\ \left. + \bra{x} C \gamma^\mu \gamma^5 C^{-1} \frac{1}{i\slashed{D}^T - m} \ket{x} \right) = \avg{\olsi{\psi} \gamma^\mu \gamma^5 \psi},
\end{gathered} 
\end{equation}
where we have also utilized Eqs.~\eqref{eq:C_prop} and \eqref{eq:transposeD} in our derivation.

Regarding the minimal subtraction term, we have:
\begin{equation}
\delta \left(\Delta_M R\right) = \delta \left(\Delta_M \right) R + \Delta_M \delta R, 
\end{equation}
with $\delta \Delta_M = 2 \eta^2 S^\mu (\delta S_\mu)/M^2$, and: 
\begin{equation}
\label{eq:R}
\hspace{-7pt}
\scalemath{0.75}{\Delta_M \delta R = 2 \eta^2 \left[\frac{1}{M^2} S^\alpha (\partial_\nu S_\alpha) S^{\mu \nu} + \Delta_M \left(\partial_\nu S^{\mu \nu} - 8 m^2 S^\mu\right) \right] \delta S_\mu}, 
\end{equation}
Under the quasi-static assumption, it follows that $\partial_\mu S^2 \approx \partial_\mu S^{\mu \nu} \approx 0$. Consequently, we find that $\avg{j^\mu_5}_{\text{eff}}$ can be expressed as:\\
\vspace{-35pt}
\begin{strip}
\raggedright
%\vspace{-10pt} % Center the line
\rule{8cm}{0.5pt}%[0.5em]{8cm}{0.5pt}
%\rule{\textwidth}{0.5pt} % Horizontal line spanning the text width with a thickness of 0.5pt
%\vspace{1em}
\begin{equation}
	\label{eq:chiral_curr}
	\hspace{-8pt}
	\begin{aligned}
		\scalemath{0.98}{\avg{j^\mu_5}_{\text{eff}} = \avg{\olsi{\psi} \gamma^\mu \gamma^5 \psi} - \frac{\eta}{24 M^2 \pi^2} \left(e^2 F^2_{\alpha \beta} + \eta^2 S^2_{\alpha \beta} - 8 \eta^2 m^2 S^2 \right) S^\mu + \frac{\Delta_M}{3 \pi^2} \eta m^2 S^{\mu}}
	\end{aligned}, 
\end{equation}
\raggedleft
%\vspace{-15pt} \\ % Add some vertical space
\hfill \rule{8cm}{0.5pt} % Another horizontal line
\vspace{-10pt} % Add some vertical space
\end{strip}
%\vfill
%\vspace{-10pt}
The divergence of all terms, except the first, vanishes under quasi-static assumptions, and the relation $\left\{\Pi_\mu, S^\mu \right\} = 0$ implies $\partial_\mu S^\mu = 0$ for real fields. Therefore, we conclude with:
\begin{equation}
\label{eq:divj5expr}
\partial_\mu \avg{j^\mu_5}_{\text{eff}} = \partial_\mu \avg{\olsi{\psi} \gamma^\mu \gamma^5 \psi} 
\end{equation}
as expected, which ensures the consistency of our results. We now can employ the classical continuity equation $\partial_\mu \avg{j^\mu_5}_{\text{eff}} = 2 i m \avg{\bar{\psi} \gamma^5 \psi}$, where the latter term is defined as:\\
\begin{equation}
\label{eq:fulltroço}
\hspace{-10pt}
\begin{aligned}
	\avg{\bar{\psi} \gamma^5 \psi} & = - \frac{m}{16 \pi^2} \int_0 ^\infty d\tau \ \frac{e^{-\tau M^2} \mathcal{G}}{\Im \cosh{\left(\tau \overline{\mathcal{X}}_{+} \right)}}\ \times \\ & \times \mathrm{tr} \left[ \gamma^5 e^{\tau \left(2 m \eta \slashed{S} \gamma^5 - \frac{1}{2} V_{\mu \nu} \sigma^{\mu \nu} \right)} \right].
\end{aligned} 
\end{equation}

A rigorous treatment of \eqref{eq:fulltroço} under the conditions $S_{\mu \nu} F^{\mu \nu} = \tilde{S}^{\mu \nu} F_{\mu \nu} = \abs{\tilde{V}^{\mu \nu} S_\mu} = 0$, $\mathcal{V} \geq 2 \eta m S^2_\mu$ and $m^2 \geq \eta^2 S^2$ reveals that the trace becomes a hyperbolic cosine whose argument is the shifted function $\tau (\overline{\mathcal{X}}_+ - \delta)$, where $\delta \approx \eta^2 m^2 S^2 / \overline{\mathcal{X}}_+$ is a perturbation. In this case, we obtain the exact ratio:
\vspace{-20pt}
\begin{strip}
\raggedright
%\vspace{-25pt}\\ % Center the line
\rule{8cm}{0.5pt}%[0.5em]{8cm}{0.5pt}
%\rule{\textwidth}{0.5pt} % Horizontal line spanning the text width with a thickness of 0.5pt
%\vspace{1em}
\begin{equation}
	\label{eq:finestruct1}
	%\hspace{-8pt}
	%\begin{gathered}
	\scalemath{0.9}{\frac{\mathrm{tr} \left[ \gamma^5 e^{\tau \left(2 m \eta \slashed{S} \gamma^5 - \frac{1}{2} V_{\mu \nu} \sigma^{\mu \nu} \right)} \right]}{\Im \cosh \left(\tau \overline{\mathcal{X}}_+ \right)} = \Re \cosh (\tau \delta) + \cot (\tau E) \Im \sinh (\tau \delta) - \coth (\tau B) \Re \sinh (\tau \delta) + \frac{\Re \cosh \left(\tau \overline{\mathcal{X}}_+ \right)}{\Im \cosh \left(\tau \overline{\mathcal{X}}_+ \right)} \Im \cosh (\tau \delta)}, 
	%\end{gathered}
\end{equation}
\raggedleft
%\vspace{-15pt} \\ % Add some vertical space
\hfill \rule{8cm}{0.5pt}  % Another horizontal line
\vspace{-10pt} % Add some vertical space
\end{strip}
\vspace{-10pt}
\\where we used the identities $\Re \sinh{(\tau \overline{\mathcal{X}}_+)} = \sinh{(\tau B)} \cos{(\tau E)}$, $\Im \sinh{(\tau \overline{\mathcal{X}}_+)} = \cosh{(\tau B)} \sin{(\tau E)}$ and $\Im \cosh{(\tau \overline{\mathcal{X}}_+)} = \sinh{(\tau B)} \sin{(\tau E)}$. This exact expression suggests a non-linear modulation of the axial current divergence by the $S_\mu$ field. For small $\delta$, this complex structure simplifies to the first-order correction:
\begin{equation}
\label{eq:finestruct2}
\hspace{-8pt}	
\begin{gathered}
	\scalemath{0.83}{\frac{\Im \cosh \left[\tau \left(\overline{\mathcal{X}}_+ - \delta\right) \right]}{\Im \cosh \left(\tau \overline{\mathcal{X}}_+ \right)} \approx 1 - \frac{\Im \left[ \tau \delta \sinh(\tau \overline{\mathcal{X}}_+) \right]}{\Im \cosh \left(\tau \overline{\mathcal{X}}_+ \right)} + \mathcal{O}(\delta^2)}.
\end{gathered} 
\end{equation}
One can see that this ``fine structure'' surfaces only when the axial intensity $\eta S$ begins to compete with the fermion's rest mass. In the UV-limit ($\tau \to 0$) we recover the covariant anomaly:
\begin{equation}
\label{eq:j5nonreg}
\begin{aligned}
	\partial_\mu \avg{j^\mu_5}_{\text{eff}} & \approx \frac{m^2}{2 \pi^2} \int_0 ^\infty d\tau \, e^{-\tau M^2} \mathcal{G} \\ & \approx \frac{1}{8 \pi^2} \left(e^2 \tilde{F}^{\mu \nu} F_{\mu \nu} + \eta^2 \tilde{S}^{\mu \nu} S_{\mu \nu}\right).
\end{aligned} 
\end{equation}

A comparative analysis between the high-energy asymptotic limit obtained here, \eqref{eq:j5nonreg}, and the standard perturbative results, Eqs. (\ref{eq:diaga}) and (\ref{eq:diagb}), reveals a discrepancy characterized by a local term proportional to $\tilde{S}^{\mu \nu} S_{\mu \nu}$, specifically differing by a factor of $2/3$. This distinction is deeply rooted in the fundamental nature of the regularization schemes employed. In standard perturbative triangle diagrams, bosonic symmetry is explicitly enforced on the external momentum vertices to fix inherent routing ambiguities. Conversely, the proper-time approach employed in our exact effective action framework treats the background tensors as continuous, macroscopic classical fields from the outset. The global regularization provided by the proper-time integration regulates the functional determinant without artificially imposing point-like permutation symmetries on the classical background. Such scheme-dependent shifts in local polynomials of the background fields are well-documented phenomena when confronting perturbative vertex expansions with exact functional methods. Therefore, the overall structure is fully consistent with the general expectations of effective field theories, rendering a direct algebraic matching via Wess-Zumino consistency conditions unnecessary for the physical dynamical regime considered here.

\subsection{The Chiral Current Divergence Beyond the Quasi-Static Limit}

The quasi-static approximation, while analytically robust, inherently suppresses the localized propagation dynamics of the axial field. To fully encompass scenarios beyond well-behaved subspaces, where the pseudo-vector background exhibits significant spacetime variation, we must relax the conditions $\{\Pi_\mu, S^\mu\} = 0$ and $\partial_\mu S^{\mu \nu} = 0$. In this subsection, we extend the computation of the chiral current divergence to the general dynamical regime, by semi-classically treating the axial background as a propagating Proca field that self-consistently responds to the induced effective current, according to the equations of motion from \eqref{eq:Proca_act} and \eqref{eq:Dirac_act}, such that $\partial_\nu S^{\nu\mu} + M_S^2 S^\mu = \eta \langle j_5^\mu \rangle$. By substituting this relation into the expression in the contribution from \eqref{eq:R} for the effective current and isolating $\langle j_5^\mu \rangle$, one naturally arrives at:
\\
%\vspace{-15pt}
%\begin{strip}
%	\raggedright
%\vspace{-5pt} % Center the line
%	\rule{8cm}{0.5pt}%[0.5em]{8cm}{0.5pt}
%\rule{\textwidth}{0.5pt} % Horizontal line spanning the text width with a thickness of 0.5pt
%\vspace{0pt}
\begin{equation}
\label{eq:full_5curr}
%\hspace{-4pt}
\begin{aligned}
	\avg{j^\mu_5}_{\text{\text{Gen}}} & = \mathcal{N} \biggr\{ \avg{j^\mu_5}_{\text{eff}} + \frac{\eta}{24 \pi^2} \biggr[M^2_S S^\mu - \\ & \left. \left. - \frac{1}{M^2} S^\alpha \left(\partial_\nu S_\alpha \right) S^{\mu \nu} \right]\right\},
\end{aligned}
\end{equation}
%\raggedleft
%\vspace{-15pt}\\ % Add some vertical space
%\hfill \rule{8cm}{0.5pt} % Another horizontal line
%\vspace{-10pt} % Add some vertical space
%\end{strip}
%\\

\noindent
Here, $\avg{j^\mu_5}_{\text{\text{Gen}}}$ is the effective current of the generalized case, while $\avg{j^\mu_5}_{\text{eff}}$ retains the definition given in Eq.~\eqref{eq:chiral_curr}. The normalization factor $\mathcal{N} = 24 \pi^2/(24 \pi^2 + \eta^2 \Delta_M)$ emerges from the algebraic isolation of the chiral current when the Proca field's back-reaction is taken into account. It is important to remark that in the previous sections, where the quasi-static approximation $\partial_\mu S^{\mu \nu} \approx 0$ was employed, this normalization factor was effectively unity. Outside the strict validity of the quasi-static limit, the effective action serves as a primary approximation while the Proca dynamics introduces this corrective normalization to ensure self-consistency between the field and its induced source. Finally, the divergence of the chiral current results in:
\vspace{-15pt}
\begin{strip}
\raggedright
%\vspace{-5pt} % Center the line
\rule{8cm}{0.5pt}%[0.5em]{8cm}{0.5pt}
%\rule{\textwidth}{0.5pt} % Horizontal line spanning the text width with a thickness of 0.5pt
%\vspace{0pt}
\begin{equation}
	\label{eq:full_divj5}
	\begin{aligned}
		\scalemath{0.98}{\partial_\mu \avg{j^\mu_5}_{\text{\text{Gen}}}} & = \scalemath{0.98}{\mathcal{N} \left[\partial_\mu \avg{\olsi{\psi} \gamma^\mu \gamma^5 \psi} - \frac{\eta^2 \left(1 + 4 \mathcal{N} \right)}{48 \pi^2 M^2} \avg{j^\mu_5}_{\text{\text{Gen}}} \partial_\mu S^2 + \frac{\eta}{24 \pi^2} \left(M^2_S + 8 m^2 \Delta_M - \frac{1}{M^2} R\right) \left(\partial_\mu S^\mu \right) + \right.} \\ & \scalemath{0.98}{\left. + \frac{\eta}{48 \pi^2 M^2} \left(M^2_S - 2 \frac{\eta^2}{M^2} R + 32 \eta^2 m^2 \right) S^{\mu} \partial_\mu S^2 - \frac{\eta}{24 M^2 \pi^2} S^\mu \partial_\mu \left(e^2 F^2_{\alpha \beta} + \eta^2 S^2_{\alpha \beta} \right) \right]},
	\end{aligned}
\end{equation}
%\raggedleft
\vspace{-20pt}\\ % Add some vertical space
\hfill \rule{8cm}{0.5pt} % Another horizontal line
\vspace{-15pt}\\ % Add some vertical space
\end{strip}
%\vfill \break
%\vspace{-10pt}
\noindent
where $\partial_\mu \avg{\olsi{\psi} \gamma^\mu \gamma^5 \psi}$ represents the expression obtained in Eqs. (\ref{eq:divj5expr}) and (\ref{eq:fulltroço}), which encompasses both the topological anomaly core [Eq. \eqref{eq:j5nonreg}] and the contributions from the vacuum back-reaction [Eq. \eqref{eq:finestruct1}].

This equation highlights the contributions to the effective current that can be anticipated across a variety of contexts involving varying fields, reinforcing the notion that the result remains valid as a first approximation. Notably, the Minimal Subtraction terms $\Delta_M R$ give rise to contributions similar to the Nieh--Yan's term \cite{TorsionNiehYan} ($\sim S^\mu \partial_\mu S^2$ in flat contorted spacetime), along with additional terms, some of which present the regularization term $\varepsilon$ from $\Delta_M$. This indicates that these contributions depend on the scale and context of the studied system: while the renormalization scheme may introduce problematic ambiguities in fundamental scenarios, such as theories of quantum gravity with torsion, the situation shifts when focusing on condensed matter physics. In this latter case, the lattice provides a physical cutoff in Weyl semimetals, effectively replacing the ambiguity associated with the regulator by the scale of the Brillouin zone~\cite{niehYansemimetals}. It is worth noting that while the topological core of the divergence remains scale-invariant, the dynamical contributions proportional to $\Delta_M$ are subject to standard renormalization. The dependence on the scale $\mu$ in these terms reflects the physical vacuum polarization in the presence of dynamical backgrounds, distinguishing them from the purely anomalous residue. When the invariants are fixed, most terms in Eq.~\eqref{eq:full_divj5} vanish, simplifying the expression to: 
\begin{equation}
\label{eq:divj5invar}
\begin{gathered}
	\partial_\mu \avg{j^\mu_5}_{\text{Static}} = \mathcal{N} \biggr[\partial_\mu \avg{\olsi{\psi} \gamma^\mu \gamma^5 \psi} + \\ + \frac{\eta}{24 \pi^2} \left(M^2_S + 8 m^2 \Delta_M - \frac{1}{M^2} R\right) \left(\partial_\mu S^\mu \right) \biggr],
\end{gathered} 
\end{equation}
which can be achieved by real fields when $\left\{\Pi_\mu, S^\mu \right\} \neq 0$.

In any case, a non-vanishing divergence of the chiral current indicates a violation in the conservation of particle number associated with each chirality, i.e., from \eqref{eq:winding} and the conservation of vector current, it follows that:
\begin{equation}
\label{eq:Nviolation}
\begin{gathered}
	\int d^4 x \partial_\mu \avg{j^\mu}_{\text{eff}} = 0 \implies \Delta N_R =-\Delta N_L,
\end{gathered} 
\end{equation}
which is a crucial requirement for scenarios such as baryogenesis \cite{Rubakov}. We will leverage this insight in the subsequent section to explore particle creation and develop a toy model of pseudo-vector-driven baryogenesis as potential cause for the observed baryonic asymmetry in the contemporary universe.

Beyond the rigid topological breaking dictated by the purely anomalous terms, the emergence of these background-dependent contributions in the chiral divergence encapsulates the phenomenon of dynamical chiral symmetry breaking (DCSB) \cite{NambuJona-Lasinio,DSB}. In this non-perturbative regime, the vacuum acts as a polarizable medium where the continuous creation and annihilation of virtual pairs under the influence of macroscopic fields dynamically sculpts the chiral flow. This aligns with the principles of DCSB, where vacuum restructuring generates macroscopic physical consequences, akin to the generation of effective masses in phenomenological models.% By capturing the complete back-reaction of the Proca field, our effective action approach demonstrates that chiral non-conservation in realistic macroscopic scenarios is profoundly dynamical, moving beyond strict topological constraints to include the fully dressed response of the quantum vacuum.

\break

\section{Phenomenological Applications\label{sec:Apps}}

In this section, we exploit the non-unitarity of the effective action to derive pair production rates and analyze a model of baryogenesis as a proof-of-concept. Furthermore, we will discuss potential extensions and implications of our findings in the realms of condensed matter physics and other related fields.\\
%\vspace{-30pt}

\subsection{Pair Production Rates\label{subsec:pairprod}}

Let us begin by defining the Vacuum Transition Probability as $P_{\text{vac}} = |\bra{A, S}\mathcal{S}_{\text{eff}}\ket{A, S}|^2$, where $\mathcal{S}_{\text{eff}} = \exp{(-i W [A, S])}$ is the S-matrix constructed from the full effective action. This probability quantifies the likelihood of remaining within the same background field configuration following scattering processes. According to the framework established by Schwinger \cite{schwinger1951}, we observe that the quantity:
%\vspace{-5pt}
\begin{equation}
%\vspace{-15pt}
\hspace{-4pt}
\scalemath{0.95}{P_{\text{creation}} \left[A, S \right] = 1-P_{\text{vac}} = 1-\exp{\left(-2 \Im W [A,S]\right)}} 
\end{equation}
%\vspace{-20pt}
provides a measure of the probability of exiting this vacuum state, thereby enabling the creation of any number of particles, therefore moving outside the scope of the effective action. This phenomenon can be modeled as a Poisson process \cite{pairPosson}, where the pair creation rate per unit volume and time is given by $\rho [A, S] =-2 \Im \mathcal{L}$.

Although a nonzero vacuum transition probability may always exist, it becomes significant primarily in field configurations where the poles of $\mathcal{L}$ are dominant. For both classes of effective actions identified, these poles are related to the generalized analog of the electric field in a manner similar to the original EH action. Specifically, they occur at $\tau_k = \pi k /E_I^{(H)}$.

In the context of first-class Lagrangians, assuming $B =-i E$, the particle creation rates associated with each sector in ASD-b subspace, for instance, can typically be expressed as $\rho_{\mathrm{ASD-b}}^{(R)} = 0$ and:

\vfill

\break

%\vspace{-20pt}
\begin{strip}
\raggedright
\vspace{-25pt}\\ % Center the line
%\rule{8cm}{0.5pt}%[0.5em]{8cm}{0.5pt}
%\rule{\textwidth}{0.5pt} % Horizontal line spanning the text width with a thickness of 0.5pt
%\vspace{-10pt}
\begin{equation}
	\label{eq:rate1L}
	\begin{aligned}
		\scalemath{1.0}{\rho^{(L)}_{\mathrm{ASD-b}}} & \scalemath{0.93}{= \frac{E^{(L)}_V}{8 \pi^2} \left\{\frac{E_V^{(L)}}{\pi} \sum_{k=1}^{\infty} \frac{1}{k^2} \cosh{\left(\frac{4 \pi k}{E^{(L)}_V} \sqrt{\mathcal{V}} \right)} e^{-\frac{\pi k m^2}{E_V ^{(L)}}}-m^2 \ln{\left[1-\cosh{{\left(\frac{4 \pi}{E^{(L)}_V} \sqrt{\mathcal{V}}\right)} e^{-\frac{\pi m^2}{E_V ^{(L)}}}}\right]}-\right.} \\ & \scalemath{1.0}{\left.-4 \sqrt{\mathcal{V}} \ln{\left[1-\sinh{\left(\frac{4 \pi}{E^{(L)}_V} \sqrt{\mathcal{V}}\right)} e^{-\frac{\pi m^2}{E_V ^{(L)}}} \right]} \right\}}.
	\end{aligned} 
\end{equation}
\raggedleft
\vspace{-10pt}\\ % Add some vertical space
\hfill\rule{8cm}{0.5pt} % Another horizontal line
\vspace{-10pt} % Add some vertical space
\end{strip}
\vspace{-10pt}
%\\
%\vspace{-10pt}

It is important to note that in such configuration, the vacuum state of right-handed fermions is stabilized, resulting in the absence of particle creation for this chirality. In contrast, for left-handed fermions, we observe an inevitable pair production phenomenon induced by a strong generalized electromagnetic field, irrespective of the value of $\mathcal{V}$.
%\vspace{15pt}
Remarkably, Eq. \eqref{eq:rate1L} also demonstrates that the pair production rate of left-handed fermions can be effectively enhanced by adjusting the parameter $\mathcal{V}$.
%\vspace{-30pt}

In the context of ASD-b scenarios, one may encounter conditions such that $\mathcal{V} = 0$ or $S^{\mu \nu} F_{\mu \nu} = 0$. These conditions imply that $\mathcal{V} = \mathcal{F}_{(L)} = - (E^{(L)}_V)^2$. Under these circumstances, Eq.~\eqref{eq:rate1} simplifies to:
%\vspace{-10pt}
\begin{equation}
\label{eq:rate1}
\hspace{-9pt}
\begin{aligned}
	\scalemath{0.78}{\rho_{\mathrm{ASD-b}}^{(L)} = \frac{E^{(L)}_V}{8 \pi^2} \left[\frac{E^{(L)}_V}{\pi} \sum_{k=1}^{\infty} \frac{1}{k^2} e^{-\frac{\pi k m^2}{E^{(L)}_V}}-m^2 \ln{\left(1-e^{-\frac{\pi m^2}{E^{(L)}_V}}\right)}\right]}.
\end{aligned} 
%\vspace{-20pt}
\end{equation}
%\vspace{15pt}
In analogous situations, such as in the NonAnom or NFC cases, one may observe a non-zero pair creation rate for both chiralities. Each chirality sector exhibits a limiting behavior akin to that described by Eq. \eqref{eq:rate1}. However, it is important to note that while both chirality sectors can generate particle pairs, the specific field strengths $E^{(R/L)}_V = e E_A \pm \eta E_S$ suggest that one sector may require comparatively weaker fields to trigger vacuum decay. Consequently, particle generation within this chirality sector can commence at lower energy levels, leading to an overall enhancement in the production of fermions with a preferred chirality. Nonetheless, it is noteworthy that in this scenario, the chirality-asymmetric instability is not as pronounced as that observed within the ASD-b subspace.
%\vspace{15pt}

It can be observed that, if achievable, these scenarios demonstrate a \emph{chirality-asymmetric stability}, where parity is violated in such a way that one chirality sector remains ``safe,'' while the opposing sector consistently decays into particles\footnotemark[8]. This interaction leads to a phenomenon of \emph{chirality-asymmetric pair creation}, marked by a net generation of fermions favoring a specific chirality. %{Should any process propel $\eta m S^2$ beyond the residual term, the system will enter a state of complete instability, triggering rapid vacuum decay.}
%\pagebreak

Conversely, for the second class, we derive:
\vspace{-20pt}
\begin{strip}
\raggedright
%\vspace{-20pt}\\ % Center the line
\rule{8cm}{0.5pt}%[0.5em]{8cm}{0.5pt}
%\rule{\textwidth}{0.5pt} % Horizontal line spanning the text width with a thickness of 0.5pt
%\vspace{1em}
\begin{equation}
	\label{eq:rate2}
	\rho_2 \left[E, B, S\right] = \frac{E B}{4 \pi^2} \sum_{k = 1} ^{\infty} \frac{(-1)^k e^{-\frac{\pi k M^2}{E}}}{k \sinh{\left(\frac{\pi k B}{e E}\right)}} \Re \cos{\left[\pi k \sqrt{1-2 i \frac{B}{E} + \frac{\left(4 \eta^2 m^2 S^2-B^2\right)}{E^2}}\right]} 
\end{equation}
\raggedleft
%\vspace{1em} % Add some vertical space
\hfill \rule{8cm}{0.5pt} % Another horizontal line
\vspace{-10pt}\\ % Add some vertical space
\end{strip}
%\vspace{-10pt}
\vfill
\noindent
for the condition $\eta S \leq m$. This expression simplifies when $S_\mu$ is spacelike ($S^2 < 0$, e.g., when it is constant and points at the $z$ direction, as originally obtained by
%\vspace{20pt}\\
Maroto (1999) \cite{maroto99}), in the regime where $B = E_S = 0$ and $2 \eta m S \leq E$, yielding:\\
%\\
%\vspace{20pt}
\begin{equation}
\label{eq:maroto}
\hspace{-6pt}
\scalemath{0.8}{\rho_{\text{M}} = \frac{e^2 E_A^2}{4 \pi^3} \sum_{k = 1}^{\infty} \frac{(-1)^k}{k^2} \cos{\left(\pi k \sqrt{1-\frac{4 \eta^2 m^2 S^2}{e^2 E_A^2}}\right)} \ e^{-\frac{\pi k M^2}{e E_A}}}, 
%\vspace{-10pt}
\end{equation}
%\vspace{-20pt}
as expected. Notably, in the limits where $B^2 \ll E^2$ and spacelike axial vector with $4 \eta^2 m^2 S^2 \ll E^2$, we can reformulate Eq. \eqref{eq:rate2} as:\\
\vspace{-30pt}
\begin{strip}
\raggedright
%\vspace{-30pt}\\ % Center the line
\rule{8cm}{0.5pt}%[0.5em]{8cm}{0.5pt}
%\rule{\textwidth}{0.5pt} % Horizontal line spanning the text width with a thickness of 0.5pt
%\vspace{1em}
\begin{equation}
	\label{eq:rate2b}
	\rho_2 \left[E, B, S\right] = \frac{E B}{4 \pi^2} \sum_{k = 1} ^{\infty} \frac{(-1)^k}{k} \coth{\left(\pi k \frac{B}{E}\right)} \cos{\left\{\pi k \left[1-\frac{\left(B^2 + 4 \eta^2 m^2 S^2\right)}{2 E^2}\right] \right\}} e^{-\frac{\pi k M^2}{E}}. 
\end{equation}
\raggedleft
%\vspace{1em} % Add some vertical space
\hfill \rule{8cm}{0.5pt} % Another horizontal line
\vspace{-10pt} % Add some vertical space
\end{strip}
\vspace{-10pt}\\
When $B^2/E^2 \approx \eta^2 m^2 S^2/E^2 \approx 0$, it can be seen that Eq.~\eqref{eq:rate2b} reproduces the results originally presented by Kim and Page (2006)~\cite{KimPagePairProd2006}.

\footnotetext[8]{This safety can be put into question if higher loops, interactions, and backreactions are included according to more complicated scenarios.}

\begin{figure*}[h!]%[20]{r}{0.3\linewidth}
\centering
\includegraphics[scale=0.3]{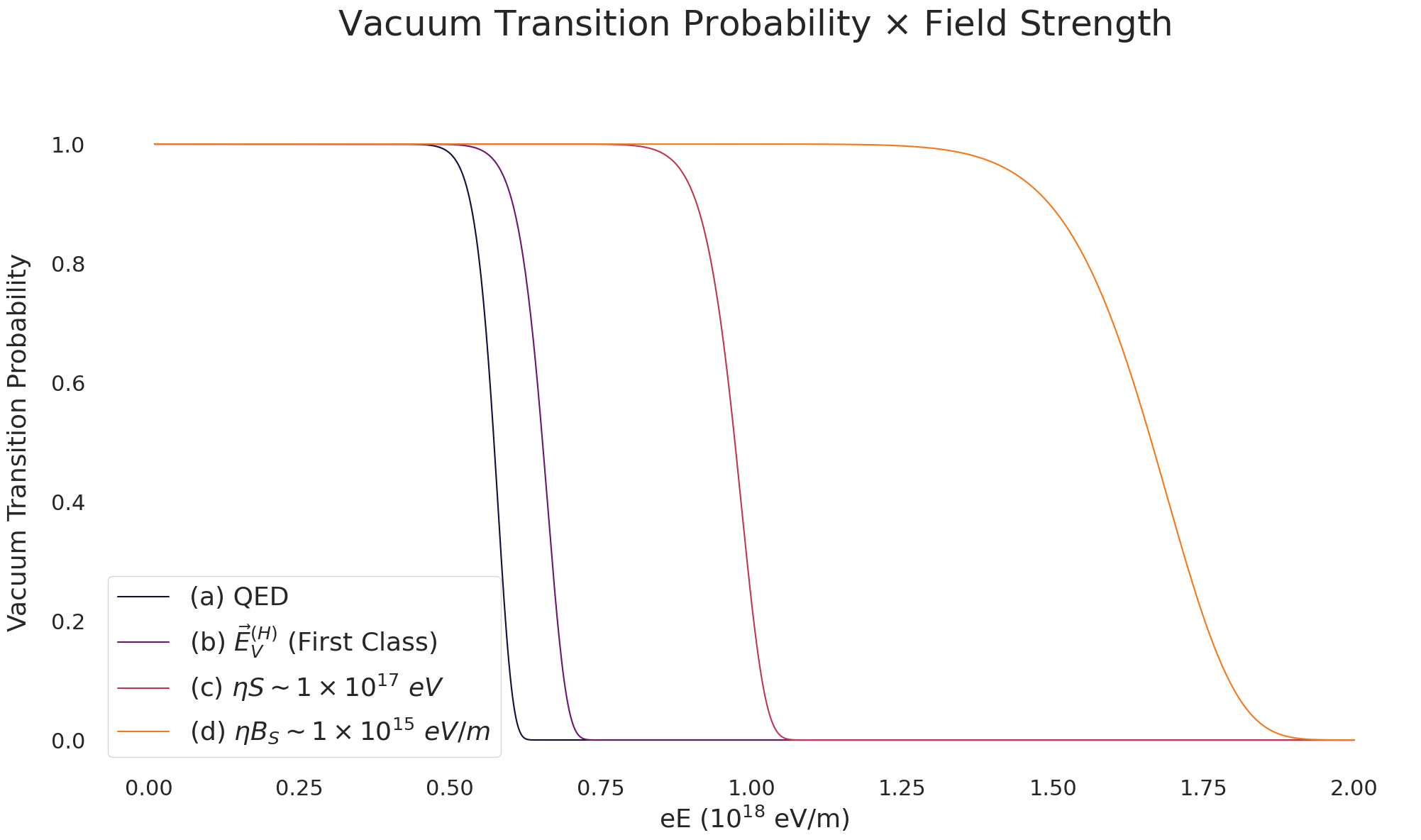}
%\captionsetup{justification=justified, format=plain}
\vspace{5pt}
\caption{\justifying
	Vacuum transition probabilities plotted via Python code~\cite{schwingermine} for an adiabatic laser pulse profile as described by Alkofer et. al. \cite{laser}, with pulse duration $\tau_{\rm pulse} = 80 \, fs$, and spatial extent $\sim \lambda = 0.15 \, nm$, according to (a) standard QED \cite{schwinger1951}, (b) Eq. \eqref{eq:rate1} for complexified fields, (c) Eq. \eqref{eq:maroto} with $\eta S \sim 1 \times 10^{17} \, e V$, and (d) Eq. \eqref{eq:rate2b} with $\eta B_S \sim 1 \times 10^{15} \, e V/m$. Note the pronounced vacuum stabilization when the axial vector is present, indicating that pseudo-vector backgrounds suppress Schwinger pair production.}
\label{fig:probs}
\vspace{-15pt}
\end{figure*}

This formulation opens avenues for exploring scenarios such as the condition where $E_S = B_A = 0$, $E_A \gg B_S$, and $E_A \gg 2 \eta m S$, among other configurations. The corresponding vacuum transition probabilities for these rates, when contrasted with Schwinger's original rate for QED, can be illustrated in FIG. \ref{fig:probs}. It is noteworthy that when present, the axial vector tends to further stabilize the vacuum. When the axial vector is time-like, it opposes the effect of the magnetic field, whereas a spacelike pseudo-vector might potentially help destabilize the vacuum under the condition $B^2 + 4 \eta^2 m^2 S^2 \geq 1$ and by diminishing the effective mass $M$, which may lead to enhancement in particle creation rates.

Additionally, we can infer that a dynamical axial vector can create particles on its own. However, this effect is invariably accompanied by the term $S^2$, unless the axial vector is massless, indicating that its pair creation rates cannot perfectly replicate those of QED if it possesses mass.

%\begin{strip}
%\raggedright
%\vspace{-25pt}\\ % Center the line
%\rule{8cm}{0.5pt}%[0.5em]{8cm}{0.5pt}
%\rule{\textwidth}{0.5pt} % Horizontal line spanning the text width with a thickness of 0.5pt
%\vspace{1em}
%\end{strip}

\subsection{Baryogenesis and Cosmology\label{subsec:cosmology}}

Particle creation, closely associated with a non vanishing chiral current divergence \cite{jnavarrosalas}, also plays a pivotal role in providing CP violations; thus, such theories align with Sakharov's conditions for baryogenesis. To delve into the feasibility of this scenario, we propose a toy model as a proof-of-concept.

Consider the presence of an axial vector (for instance, the pseudo-trace of torsion) with mass $M_S \gg m$ in the primordial universe. Assuming thermal equilibrium, once the temperature of the universe falls below $T \sim M_S$, the excitation of its new modes begins to decrease as $\propto e^{-M_S/T}$. This change allows us to model it as oscillating classically, with an energy density given by $\mathcal{E} \approx M_S + 3 T/2$.

At this juncture, the energy stored in this pseudo-vector is sufficiently substantial to initiate a particle creation process. This process will continue until the strength of its electric field analogue drops below a critical value or until it ``freezes.'' Subsequently, the remaining energy will undergo dilution until the present day. The asymmetry in particle numbers for each chirality at the conclusion of this process translates into a baryon asymmetry that remains conserved until the values recently measured \cite{PlanckColab}.

Let us assume an appropriate coarse-graining process that preserves the Markovian approximation \cite{noBHfromlight}, allowing the particle creation rates to remain approximately constant at each step. From Eq. \eqref{eq:winding}, it follows that $n_{5,0} = \int_{t_i}^{t_f} \rho [A, S] dt = N_5$  $(V_{\text{comoving}})^{-1} $ represents the total number of particles generated per co-moving volume at zero temperature, where $t_i$ and $t_f$ denote the initial and final instants of the process. At this time, the universe was dominated by radiation \cite{earlyuniverse,ParticleDataGroup}, leading to:
\begin{equation}
\label{eq:t}
\scalemath{0.95}{t = \frac{0.301 M_{\text{Planck}}}{\sqrt{g_{\ast}} T^2} \implies dt \approx-\frac{0.602 M_{\text{Planck}}}{\sqrt{g_{\ast}} T^3} dT}, 
\end{equation}
where $M_{\text{Planck}} = 1.22 \times 10^{19} \, GeV$ and, by considering all the degrees of freedom in the standard model, $g_{\ast} \approx 106.75$ is taken as a constant. Assuming the entire process occurs prior to the electroweak symmetry breaking or even before neutrinos acquire mass (alternatively, we could consider calculating the temperature correction for massless fermions as a first approximation, given that their masses are small compared to $M_S$), we follow the prescription outlined by Kim, Lee and Yoon \cite{particlecreationtemp}. This involves assuming maximum efficiency in particle creation---thus using QED's rate as an approximation---yielding:
\begin{equation}
\label{eq:N5}
\begin{aligned}
	n_{5} & = \frac{1}{(2 \pi)^3}\int n_{5,0} \tanh{\left(\frac{\omega}{2 T} \right)} d^3 \vec{k}, \\
	& =-\frac{-0.301 \zeta \left(3\right) \eta^2 E^2_S M_{\text{Planck}}}{\left(4 \pi \right)^4 \sqrt{g_{\ast}}} \Delta T,
\end{aligned} 
\end{equation}
where $\Delta T = T_f-T_i$ represents the temperature difference between the onset and conclusion of the process.\\

We can then leverage the fact that $n_5/s = \text{constant}$ to extrapolate the measured baryon-to-photon ratio $\eta_B = n_5/n_\gamma \approx 6.12 \times 10^{-10}$, where $s \approx 7.04 n_\gamma$ today, but was $s \approx (2\pi^2/45) g_{\ast} \times$ $\times T_f^3$ during the radiation-dominated epoch. This allows us to find $n_5/s \approx 8.6932 \times 10^{-11} \sim 10^{-10}$ in the past, leading to $n_5 \approx (2\pi^2 g_{\ast} T_f^3/45) (8.6932 \times 10^{-11})$. Furthermore, we assume that energy storage oscillates between the invariant energy densities $M^2_S \langle S^2 \rangle$, $E^2_S$ and $B^2_S$, sharing it evenly on average as follows: $ \mathcal{E}/3 = M^2_S \langle S^2 \rangle/2 = E^2_S/2 = B^2_S/2$. Thus:
\vspace{-20pt}\\
\begin{strip}
%\raggedright
\vspace{-20pt}\\%\\ % Center the line
\rule{8cm}{0.5pt}%[0.5em]{8cm}{0.5pt}
%\rule{\textwidth}{0.5pt} % Horizontal line spanning the text width with a thickness of 0.5pt
\vspace{-10pt}
%\vspace{1em}
\begin{equation}
	\vspace{-10pt}
	\label{eq:Spast}
	\eta \sqrt{\avg{S^2}} = \frac{2 \left(2 \pi \right)^3 T_f}{M_S \sqrt{45 M_{\text{Planck}}}} \left(\frac{T_i}{T_f}-1 \right)^{-\frac{1}{2}} \left(\frac{4 g_{\ast}^3}{\zeta^2 \left(3\right)}\right)^{\frac{1}{4}} \left(1.6974 \times 10^{-5}\right) \, GeV^{\frac{3}{2}}. 
\end{equation}
\raggedleft\\
%\vspace{-10pt}
%\vspace{1em} % Add some vertical space
%\hfill \rule{8cm}{0.5pt} % Another horizontal line
%\vspace{-10pt}
%\vspace{-5pt} % Add some vertical space
\end{strip}
%\vspace{-10pt}
Reorganizing in terms of scale as $T = \alpha M_S$, we derive:
\begin{equation}
\label{eq:Spastvalue}
\eta \avg{{S}}_{\text{rms}} \approx 1.54 \frac{\alpha_f^{3/2}}{\sqrt{\alpha_i-\alpha_f}} 10^{-11} \, GeV. 
\end{equation}
This implies that if the process concludes at temperatures in between $T_f \sim 10^{-2} M_S$ to $10^{-3} M_S$, and the temperature variation spans an order of $10^{-1}$ to $10^1$, then values ranging from $\eta \sqrt{\langle S^2 \rangle} = \eta \langle S \rangle_{\text{rms}} \sim 10^{-14} \, GeV$ to $10^{-12} \, GeV$ would suffice to account for the present baryon asymmetry. Notably, these conditions yield expected values smaller than the upper bound for the total mass of neutrinos, thereby demonstrating consistency with the requirement that $m^2-\eta^2 \langle S^2 \rangle \geq 0$ for well-behaved pair creation rates.

The density of these relics will dilute as $\propto a^{-3}$, where $a(T)$ $\propto [T^3 g_{\ast s}(T)]^{-1/3}$ is the scale factor. Thus, to obtain the relic density, we can perform:
\begin{equation}
\label{eq:comparison}
\frac{\avg{\mathcal{E}_{\text{relic}}}}{\avg{\mathcal{E} (T_f)}} = \frac{g_{\ast s} (T_{\text{today}})}{g_{\ast s} (T_f)} \left[\frac{T_{\text{today}}}{T_f} \right]^3 = \frac{\avg{S^2_{\text{relic}}}}{\avg{S^2}_{T_f}}, 
\end{equation}
given $T_{\text{today}} = 2.73 \, K = 2.3 \times 10^{-13} \, GeV$ and $g_{\ast s}(T_{\text{today}}) = 3.91$. We conclude that: 
\begin{equation}
\label{eq:Stoday}
\scalemath{0.96}{\eta \avg{S_{\text{relic}}}_{\text{rms}} \lessapprox \frac{6.318}{m_S \sqrt{m_S \left(\alpha_i-\alpha_f\right)}} \times 10^{-61} \, GeV}, 
\end{equation}
where we denote the mass in Planck units at the conclusion, i.e., $M_S = m_s M_{\text{Planck}}$. Ultimately, from $M^2_S \langle S^2 \rangle = 2 \mathcal{E}/3$, the relic energy density is approximately:
\begin{equation}
\label{eq:relicenergy}
\mathcal{E}_{\text{relic}} \lessapprox \frac{6.8676}{\eta^2 m_S \left(\alpha_i-\alpha_f\right)} 10^{-83} \, GeV^4. 
\end{equation} 
This relic energy density is remarkably low, comfortably situated well below current observational limits on Lorentz-violating interactions from astrophysical sources (e.g., $\gamma$-ray polarimetry) and laboratory tests~\cite{Lorentzlimits}. The corresponding value of the root-mean-square axial field strength $\eta \langle S_{\text{relic}}\rangle_{\text{rms}} \lesssim 10^{-61}$ ${\rm GeV}$ is also consistent with precision tests of CPT invariance~\cite{Kosteleck_2011}. This parametric suppression arises naturally from the dilution factor $\sim (T_{\text{today}}/T_f)^3$ combined with the stipulation that baryogenesis occurs before the electroweak symmetry breaking. In summary, despite the simplicity of this model, more rigorous constructions of such scenarios could indeed be viable explorations for future research.

It is noteworthy that the relic energy can be finely tuned to approximately $10^{-47} \, GeV^4$, while maintaining $\eta \langle S \rangle_{\text{rms}}$ below the bounds imposed by experimental evidence. This observation indicates that such axial vector may also serve as a viable candidate for Dark Energy, given that massive vector fields can exhibit negative pressure \cite{DarkProca}. In fact, the subset of parameter space where $\left\{\Pi_\mu, S^\mu \right\} \neq 0$ could present intriguing possibilities for such theoretical proposals \cite{H5notzero,DarkProcaConstraints}. Nonetheless, it is essential to conduct a careful stability analysis to evaluate the implications of this hypothesis rigorously.

\subsection{Weyl Semimetals, Quark-Gluon Plasma, Weak Force and Beyond\label{subsec:other}}

The electroaxial effective action derived above has direct applications to time-reversal asymmetric Weyl and Dirac semimetals, where axial-vector backgrounds emerge from Weyl-node separation $b_\mu = (b_0, \vec{b})$ in momentum space, strain-induced pseudomagnetic fields, and magnetic textures \cite{EHinDiraccondensedmatter,Diractorsion2,Diractorsion,dynamicalstrain,torsionweylmetal,EMWeylsemimetals}. In this context, the node separation vector plays the role of $S_\mu$, and our nonperturbative results become experimentally accessible: the required magnetic field strengths ($\sim 1$--$10$ T) are readily achievable in laboratory settings. Lorentz invariance is emergent rather than fundamental in these systems \cite{Lorentzviolationsemimetals2}, permitting nonzero mixed terms $S_{\mu \nu} F^{\mu \nu} = \vec{E}_A \cdot \vec{B}_S + \vec{E}_S \cdot \vec{B}_A \neq 0$ that induce apparent vector-current nonconservation, requiring Bardeen-Zumino polynomial counterterms for consistent anomaly structure.

Our framework yields several testable predictions. First, the additional terms in the chiral-current divergence [Eqs. \eqref{eq:full_divj5}]---which includes contributions beyond the standard Adler--Bell--Jackiw / Carroll--Field--Jackiw-Torsion [Eqs. \eqref{eq:diaga} and \eqref{eq:diagb}] and Nieh--Yan terms---modifies anomalous transport coefficients and should manifest as corrections to longitudinal magnetoconductance in materials such as TaAs or Cd$_3$As$_2$. Second, the pair-production rates derived in Eqs. \eqref{eq:rate1L}--\eqref{eq:rate2b} translate into chiral-charge pumping rates between Weyl nodes under strong applied fields, with the vacuum-stabilization effect (Fig. \ref{fig:probs}) predicting enhanced threshold fields compared to naive Schwinger estimates. This effect should be observable in pulsed magnetic field and Circular Photogalvanic Effect \cite{CPGE} experiments. Third, the electromagnetic-field generation via axial-vector decay [Eq. \eqref{eq:decaypref}] suggests that dynamical strain or magnetic textures in Weyl semimetals can source photon production analogous to phenomena observed in magnetized plasmas \cite{plasma_axial1,plasma_axial2}. Quantitative comparison with transport measurements and pump-probe spectroscopy will be pursued in subsequent work.

Furthermore, our derivation of the fine structure of the chiral current divergence [Eqs. \eqref{eq:finestruct1} and \eqref{eq:finestruct2}] and its generalization [Eq. \eqref{eq:full_divj5}] bears significant implications for the physics of the Quark-Gluon Plasma (QGP) produced in heavy-ion collisions \cite{ChiralTransportQCD}. While the Chiral Magnetic Effect is traditionally analyzed in the massless limit, the strong magnetic fields ($eB \sim m_{\pi}^2$) present in these collisions probe energy scales where quark masses cannot be fully neglected. Our results indicate that the interplay between the fermion mass and the axial background---which, in the electroweak sector, effectively represents the interaction with $Z$ and $W$ bosons---introduces a dispersive modulation to the topological charge pumping. Specifically, the chirality-asymmetric instability identified in the ASD sector [Eq. \eqref{eq:rate1L}] suggests that in regimes where the weak sector fields acquire classical expectations (e.g., during electroweak symmetry breaking or in localized condensates), the generation of chiral imbalance is %not merely topological but
dynamically reinforced by the vacuum decay rates, potentially altering the transport coefficients usually assumed in anomalous hydrodynamics.

Finally, we emphasize that the analytical framework developed for the ASD-and MT-Type subspaces offers %more than a formal classification; it serves as a robust first-principle 
approximations for Weak-Force processes in extreme environments, such as neutrino pair creation or $\nu-e$ scattering in dense media. While the full electroweak sector is non-Abelian, one can employ an Abelian Cartan projection for the near-field dynamics. This approximation is justified by the hierarchy of scales in the bound-state formation: for the relevant field amplitudes ($\abs{W_\mu} \sim M_W/g$), non-Abelian corrections are parametrically suppressed by $\mathcal{O}(g^2/4\pi)$ \cite{tHooft1981,KronfeldSchierholzWiese1987,EzawaIwazaki1982,Bonati2010}. Furthermore, recent real-time lattice simulations of the Schwinger mechanism confirm that Abelian dominance holds for the short formation times ($t_{\text{form}} \ll \Gamma_W^{-1}$, where $\Gamma_W$ is the weak boson width) characteristic of this resonant pair production \cite{GelisTanji2016,SpitzBerges2019,CopingerMorales2021}. This separation of time scales also validates the quasi-static approximation used to diagonalize the Hessian here, treating the background fields as locally constant during the process. Consequently, our non-perturbative effective actions and decay rates provide a good description of the ``Abelian dominance'' regime of the electroweak sector, serving as a reliable effective field theory for modeling vacuum stability and particle production in scenarios where weak bosons condense or exhibit strong classical expectations.

Beyond these avenues, two speculative extensions merit mention, though they require significant theoretical development. First, the longitudinal mode of a massive pseudo-vector $S_\mu$ could function as an axion-like coupling if the anomaly structure [Eq. \eqref{eq:anominduced}] is localized via auxiliary St\"{u}ckelberg or Higgs fields, addressing the strong-CP problem as in axion models \cite{axion2,axion3,axion4,axion5}; our sector classification (Sec. \ref{sec:hessian}) would then constrain viable parameter space. Second, in Standard Model extensions with magnetic monopoles, the anomalous $S \to \gamma\gamma$ decay channel (Sec. \ref{sec:Anom}) could catalyze monopole-antimonopole pair production \cite{rajantiemonopole} in heavy-ion collisions, with potential observational implications for the MoEDAL \cite{moedal} and RHIC \cite{rajantie2} experiments; however, realizing this scenario requires promoting $A_\mu$ to a non-Abelian gauge field, placing it outside the scope of the present Abelian treatment.

\section{Discussion and Outlook\label{sec:conclusions}}

This work presents a comprehensive one-loop analysis of electroaxial theory---the coupling of Dirac fermions to electromagnetic and massive axial-vector backgrounds---within a controlled quasi-static approximation. Through complete diagonalization of the functional Hessian, we have achieved several key advances: (i) a systematic classification of the theory's parameter space into physically distinct sectors with well-defined stability properties; (ii) exact closed-form expressions for the Euler-Heisenberg effective action across all viable regimes, extending beyond perturbative expansions; and (iii) novel nonperturbative predictions for vacuum stability, Schwinger pair production thresholds, and the structure of the chiral current divergence. Our results unify and generalize previous treatments of both pure QED and axial-coupled systems, while revealing robust vacuum-stabilization mechanism with possibility of chiral-asymmetric pair production operative across the entire physically accessible parameter space.

The framework developed here provides both conceptual clarity and practical utility. By deriving explicit criteria for vacuum stability and identifying the precise conditions under which each regime applies, we have established a complete map of the theory's one-loop structure. Our expressions reduce correctly to known limits---recovering the standard Heisenberg-Euler Lagrangian when axial fields vanish and reproducing Maroto's (1999)~\cite{maroto99} results in the appropriate parameter regime---while simultaneously extending the analysis to previously unexplored sectors of configuration space. The verification of consistency with third-order perturbative expansions, combined with the exact nature of our one-loop results, provides strong confidence in the robustness of our approach.

Several natural directions emerge for extending this analysis. First, the regions of parameter space we have identified as stable provide concrete targets for phenomenological applications. While our analysis establishes \emph{where} these regimes exist and \emph{what} their properties are, connecting these mathematical structures to specific physical systems---whether in condensed matter realizations with Weyl semimetals, Standard Model's electroweak theory and its extensions incorporating axial gauge bosons, or cosmological scenarios with torsion-induced axial fields---represents an exciting avenue for future work. Each application domain will impose its own constraints on field magnitudes, coupling strengths, and mass scales, determining which portions of our parameter space are physically realized.

Second, extending beyond the quasi-static approximation would open access to dynamical phenomena currently beyond our framework's reach. Time-dependent backgrounds, particularly those varying on timescales comparable to the axial mass, could reveal phenomena invisible in the constant-field limit. This includes contributions from topological invariants such as the Nieh-Yan term, which vanishes identically in our present analysis but may play crucial roles in evolving geometries. Such extensions would be particularly relevant for modeling ultrafast laser interactions in Weyl materials, early-universe phase transitions, or other scenarios where non-adiabatic effects dominate the vacuum response.

Third, while our one-loop results are exact within their domain of validity, a complete understanding of the theory's quantum structure requires analysis at all loop orders. A systematic Functional Renormalization Group~\cite{functionalrenormalization} study would address the interplay between our one-loop effective action and higher-order corrections, determine the theory's UV behavior, and establish its consistency with potential UV completions. This is especially pertinent when the axial field originates from torsion~\cite{Nonpertrenorm}, where compatibility with quantum gravity constraints becomes essential. Recent developments in non-perturbative renormalization techniques provide promising tools for such investigations.

Finally, incorporating gravitational degrees of freedom would enable direct application to cosmological scenarios. While our flat-spacetime treatment captures the essential quantum field theoretic structure, phenomena such as gravitational particle production~\cite{dobadoQOProfImprimiu}, backreaction on spacetime curvature, and the interplay between geometric torsion and axial-induced anomalies~\cite{dobadotorsion} require a fully covariant formulation. The exact one-loop expressions derived here could serve as a foundation for such generalizations, with the heat kernel methods we employ naturally extending to curved backgrounds~\cite{HeatKernelQG,HeatKernelQG2}.

Beyond these theoretical extensions, our results provide concrete, testable predictions. The stability criteria we derive impose sharp constraints on allowed field configurations, the pair production thresholds we calculate are directly observable in principle, and the chiral current divergence we uncover has measurable consequences for transport phenomena in systems with axial couplings. In condensed matter systems---particularly Weyl and Dirac semimetals where axial electromagnetic fields emerge naturally---our predictions for vacuum polarization effects and modified dispersion relations could be tested with current or near-future experimental techniques. In high-energy physics, electroweak phenomenology and searches for new gauge bosons with axial couplings could use our results to constrain parameter space and predict associated phenomenology. In cosmology, our framework provides quantitative tools for analyzing baryogenesis mechanisms, primordial magnetogenesis scenarios, and other early-universe processes where axial fields may have played a role. We hope that future experimental tests of these predictions, combined with theoretical extensions of the methods presented here, will further elucidate the dynamics of fermions coupled to vector and pseudo-vector potentials, thereby refining our understanding of the quantum vacuum structure in these fundamental interactions.

\backmatter

\bmhead{Supplementary information}

Computational implementations of the calculations presented in this work are provided as ancillary materials. These include a Jupyter notebook with Python codes and outputs for diagonalization of the functional Hessian~\cite{diagmine}, a Mathematica notebook with further draft calculation~\cite{MathematicaNotebook,Eigenvecs}, and a Python code for generating the vacuum probability transitions in Fig.~\ref{fig:probs}~ \cite{schwingermine}. Users can reproduce all numerical results and figures from the main text, explore additional parameter regimes, and adapt the code for related calculations. 

\bmhead{Acknowledgements}

I thank I. L. Shapiro, G. P. de Brito, J. A. Helay\"{e}l-Neto, W. Cesar e Silva, G. M. C. da Rocha, and P. M. M. de Souza for the discussions and useful comments.

\section*{Declarations}

%Some journals require declarations to be submitted in a standardised format. Please check the Instructions for Authors of the journal to which you are submitting to see if you need to complete this section. If yes, your manuscript must contain the following sections under the heading `Declarations':
%
%\begin{itemize}
%	\item Funding
%	\item Conflict of interest/Competing interests (check journal-specific guidelines for which heading to use)
%	\item Ethics approval and consent to participate
%	\item Consent for publication
%	\item Data availability 
%	\item Materials availability
%	\item Code availability 
%	\item Author contribution
%\end{itemize}
%
%\noindent
%If any of the sections are not relevant to your manuscript, please include the heading and write `Not applicable' for that section. 
%
\subsection*{Funding}
\noindent
This work was partially supported by the Conselho Nacional de Desenvolvimento Científico e Tecnológico (CNPq) under Grant No. 132181/2023-1.

\subsection*{Conflict of Interest}
The authors declare that they have no competing interests.

\subsection*{Ethics Approval}
\noindent
Not applicable.

\subsection*{Consent to Participate}
\noindent
Not applicable.

\subsection*{Consent for Publication}
\noindent
Not applicable.

\subsection*{Data Availability}
\noindent
All computational codes (Python and Mathematica) developed for this work are publicly available and have been deposited in GitHub repositories as cited in Refs.~\cite{diagmine,MathematicaNotebook,Eigenvecs,schwingermine}. The datasets generated during the current study (consisting of numerical evaluations of the analytical expressions presented) can be reproduced using the provided code. No experimental data apart from what is reported by \cite{PlanckColab} was used in this theoretical study.

\subsection*{Code Availability}
\noindent
The Jupyter implementation and Mathematica notebooks are available at the GitHub repositories cited in Refs.~\cite{diagmine,MathematicaNotebook,Eigenvecs,schwingermine}.

\subsection*{Author Contributions}
\noindent
Lucas Pereira de Souza confirms sole responsibility for the following: study conception and design, theoretical construction, analytical calculations, numerical simulations, data analysis and interpretation, and manuscript preparation.
%
%===================================================%
% For presentation purpose, we have included        %
% \bigskip command. Please ignore this.             %
%===================================================%
%\bigskip
%	Editorial Policies for:
%	
%	\bigskip\noindent
%	Springer journals and proceedings: \url{https://www.springer.com/gp/editorial-policies}
%	
%	\bigskip\noindent
%	Nature Portfolio journals: \url{https://www.nature.com/nature-research/editorial-policies}
%	
%	\bigskip\noindent
%	\textit{Scientific Reports}: \url{https://www.nature.com/srep/journal-policies/editorial-policies}
%	
%	\bigskip\noindent
%	BMC journals: \url{https://www.biomedcentral.com/getpublished/editorial-policies}
%\end{flushleft}
%\vspace{-20pt}

%\break

%\vspace{-20pt}
%\raggedright
%\vspace{-10pt} % Center the line
%\rule{8cm}{0.5pt}%[0.5em]{8cm}{0.5pt}

\begin{appendices}

\begin{strip}
	%\raggedright
	%\vspace{-10pt} % Center the line
	%\rule{8cm}{0.5pt}%[0.5em]{8cm}{0.5pt}
	%\rule{\textwidth}{0.5pt} % Horizontal line spanning the text width with a thickness of 0.5pt
	%\vspace{1em}
	\vspace{20pt}
	\section{Auxiliary Terms \label{app:solution}}
	\addcontentsline{toc}{section}{Appendix: Auxiliary Terms}

	In this appendix, we present the full expression of the terms from the characteristic equation and general solution presented in Sec. \ref{sec:hessian}. Let us begin with the terms from Eq. \eqref{eq:charac_eq}, which reads:
	\begin{equation}
		\label{eq:charac_defs}
		\hspace{-10pt}
		\begin{cases}
			\scalemath{1.0}{\mathcal{A} = \eta^2 \left[2 m^2 S^2 - \frac{1}{2}\left(\left\{\Pi_\mu, S^\mu \right\}\right)^2\right]-\mathcal{V}}, \\
			\scalemath{1.0}{\mathcal{B} = \frac{1}{2}\eta \mathcal{W} \left\{\Pi_\mu, S^\mu \right\}}, \\
			\scalemath{1.0}{\mathcal{C} =} \scalemath{0.87}{\mathcal{A}^2 + \mathcal{W}^2 - \eta^2 \left[2 \left(\left\{\Pi_\mu, S^\mu \right\}\right)^2 \mathcal{V} + 4 m^2 \left(\widetilde{V}^{\mu \nu} S_\nu\right)^2 \right]}.
		\end{cases}
		%\vspace{-10pt}
	\end{equation}
	The expression of $D_{\pm}$ in Eq. \eqref{eq:sol_gen} is:
	\begin{equation}
		\label{eq:Deep_into_chaos}
		\mathcal{D}_{\pm} = \sqrt[3]{\Delta}-2 \left(1 \pm 3\right) \mathcal{A} + \frac{4}{\sqrt[3]{\Delta}} \left[4 \mathcal{A}^2 + 6 \eta^4 m^2 S^2 \left(\left\{\Pi_\mu, S^\mu \right\} \right)^2 + 3 \mathcal{W}^2 \right],
	\end{equation}
	with discriminant defined as:
	\begin{equation}
		\label{eq:discr}
		\Delta = 2 \mathcal{K} + i\sqrt{2 \mathcal{K} + 64 \left\{2 \left( 2 \mathcal{A}^2 - \eta^2 \left\{\Pi_\mu, S^\mu \right\} \mathcal{V} \right) + 3 \mathcal{W}^2 + 6 \eta^2 m^2 \left[\eta^2 \left(\left\{\Pi_\mu, S^\mu \right\} \right) S^2 - 2 \left(\widetilde{V}^{\mu \nu} S_\nu\right)^2 \right] \right\}^3},
	\end{equation}
	where:
	\begin{equation}
		\label{eq:Khaos}
		\scalemath{1.0}{\mathcal{K} = 4 \mathcal{A}^3-9 \mathcal{A} \left[\mathcal{C} + 16 m^2 \eta^4 \left(\left\{\Pi_\mu, S^\mu \right\} \right)^2 S^2 \right] - 108 \mathcal{B}^2}.
	\end{equation}\\
	%\vspace{1em} % Add some vertical space
	%\hfill\rule{8cm}{0.5pt}\\ % Another horizontal line
	%\vspace{-10pt} % Add some vertical space
	%\vfill
\end{strip}
%\vspace{20pt}
%\vfill
%\pagebreak
%\vspace{10pt}

\section{Discrete Symmetries and Bilinears\label{app:bilinears}}
\setcounter{equation}{0}
\justifying

For definiteness we adopt the convention that $S_\mu$ couples to the axial fermion current, as can be seem at Eq. \eqref{eq:Dirac_act}, that is:
\begin{equation}
	\mathcal{L}^{\rm int}_5 = \eta S_\mu \olsi{\psi} \gamma^\mu \gamma^5 \psi,
\end{equation}
%\vspace{-10pt}
so that $S_\mu$ is C-even. The axial current is C-even due to Eq. \eqref{eq:C_prop}, that is:
\begin{equation}
	\left(\olsi{\psi} \gamma^\mu \gamma^5 \psi \right)^C = \olsi{\psi} C^{-1} (\gamma^\mu\gamma^5)^{T} C \psi = \olsi{\psi} \gamma^\mu \gamma^5\psi,
\end{equation}
using $C\gamma^\mu C^{-1} = -(\gamma^\mu)^T$ and $C \gamma^5 C^{-1} = (\gamma^5)^T$ together with the anticommutation relation $\{\gamma^\mu,\gamma^5\}=0$. Hence the interaction density $S_\mu J_5^\mu$ is C-even provided the coupling $\eta$ is a C-scalar; choosing $S_\mu$ to be C-even corresponds to preserving charge-conjugation symmetry in this sector.

Under spatial inversion $P$ an axial 4-vector flips sign on its time component and not on its spatial components:
\begin{equation}
	\scalemath{0.92}{j_5^0 (t,\mathbf{x}) \xrightarrow{P} - j_5^0 (t,-\mathbf{x}), \qquad j_5^i (t,\mathbf{x}) \xrightarrow{P} + j_5^i (t, -\mathbf{x})},
\end{equation}\\

\captionsetup[table]{font=small}
\captionof{table}{Parity (P), charge-conjugation (C) and CP parities of bilinear scalars constructed from $F_{\mu \nu}$, $S_{\mu \nu}$ and their duals, assuming $S_\mu$ is C-even.\label{tab:CP-parities}}
\begin{table}[h!]
	\vspace{-10pt}
	\centering
	\setlength{\tabcolsep}{3pt}
	\renewcommand{\arraystretch}{2}
	\begin{tabular}{|c|c|c|c|}
		\hline
		Scalar & P & C & CP \\
		\hline
		$F^{\mu \nu} F_{\mu \nu}$ & even & even & even \\
		$\tilde{F}^{\mu \nu} F_{\mu \nu}$ & odd & even & odd \\
		$S^{\mu \nu} F_{\mu \nu}$ & odd & odd & even \\
		$\tilde{S}^{\mu \nu} F_{\mu \nu}$ & even & odd & odd \\
		$S^{\mu \nu} S_{\mu \nu}$ & even & even & even \\
		$\tilde{S}^{\mu \nu} S_{\mu \nu}$ & odd & even & odd \\
		\hline
	\end{tabular}%
	%\captionof{table}{Parity (P), charge-conjugation (C) and CP parities of bilinear scalars constructed from $F_{\mu \nu}$, $S_{\mu \nu}$ and their duals, assuming $S_\mu$ is C-even.}
	%\label{tab:CP-parities}
\end{table}

\vspace{-15pt}
and the same transformation rules hold for $S^\mu$. Consequently, contractions of two objects with the same pseudo-character (both ordinary or both pseudo) are P-even, while mixed contractions of an ordinary with a pseudo object are P-odd. The parities (P), charge-conjugation parities (C) and CP parities of all independent bilinear scalars constructed from $F_{\mu \nu}$, $S_{\mu \nu}$ and their duals are summarized in Table~\ref{tab:CP-parities}, under this assumption.
%\vspace{10pt}

Finally, we note that taking $S_\mu$ to be C-even is a model choice tied to the assumed axial coupling; if instead $S_\mu$ were a vector gauge field for a $U (1)$ charge that flips under $C$, then the C/CP entries in Table~\ref{tab:CP-parities} must be modified accordingly.

\end{appendices}
\bibliography{biblio}
%\printbibliography

\end{document}